\documentclass[11pt,a4paper]{article}
\usepackage[T1]{fontenc}

\usepackage{amsmath,latexsym} 
\usepackage{mathrsfs}
\usepackage{mathtools}
\usepackage[dvipsnames,table]{xcolor} 
\usepackage{graphicx}
\usepackage{slashed}

\usepackage{pdflscape}
\usepackage{geometry}

\usepackage{tikz}
\usetikzlibrary{decorations.pathreplacing}
\usepackage[compat=1.1.0]{tikz-feynman}
\usepackage[nomessages]{fp}
\usepackage{jheppub}
\usepackage{graphicx}
\usepackage{soul}

\usepackage{youngtab}
\usepackage{ytableau}

\usepackage{changepage} 
\usepackage{braket}
\usepackage{relsize}
\usepackage{lscape} 
\usepackage{bbm} 
\newcommand{\pmatrixx}[1]{\begin{pmatrix} #1 \end{pmatrix}} 
\usepackage{pdfpages} 
\usepackage{nicematrix}
\usepackage{array,xparse,l3keys2e,expl3}
\usepackage{float}
\usepackage{multicol}
\usepackage{scalerel}
\usepackage[export]{adjustbox}

\usepackage{multirow}

\def\supON{{}}

\def\nn{\nonumber}
\def\R{\mathcal{R}} 

\def\eps{\varepsilon}

\usepackage[normalem]{ulem}

 \title{
\boldmath {Constraints on the $O(n)$ model from a negative number of flavors}
}

\author[-2,-1]{Daniele Artico}
\author[1,2]{and Jasper Roosmale Nepveu}

\affiliation[-2]{Dipartimento di Fisica, Universit\`a di Parma, Parco Area delle Scienze 7/A, 43124 Parma}
\affiliation[-1]{INFN, Gruppo di Parma}
\affiliation[1]{Department of Physics and Center for Theoretical Physics, National Taiwan University, Taipei 10617, Taiwan}
\affiliation[2]{
Leung Center for Cosmology and Particle Astrophysics, Taipei 10617, Taiwan}

\emailAdd{daniele.artico@unipr.it}
\emailAdd{jasperrn@ntu.edu.tw}

\abstract{
    Treating the number of flavors $n$ as a variable in the $O(n)$ model leads to non-perturbative constraints on the spectrum of operators. 
    Using the duality between $O(n)$ and $Sp(-n)$, we extend these relations to negative even values of $n$ and we make them explicit by decomposing operators with general flavor structure into irreducible representations.
    In perturbation theory, we exploit this structure to reveal novel degeneracies in the scaling dimensions of different operators, which persist for arbitrary $n$. This allows us to derive the two-loop anomalous dimension of any $\phi^k$-type operator from existing results without additional loop calculations.
    The same mechanism dictates patterns in renormalization group mixing matrices, yielding new non-renormalization results in a specific operator basis.
    We comment on extending this framework to relations between operator product expansion coefficients and to analogous constraints arising from evanescence under continuation in the number of spacetime dimensions.
}

\begin{document}

\maketitle

\section{Introduction}

It is common practice in physics to continue integer quantities to continuous variables variables. 
Most notably, the number of spacetime dimensions ($d$) can be continued to regulate divergences in quantum field theory (QFT)~\cite{Bollini:1972ui,tHooft:1972tcz}. Moreover, observables can often be expressed as a function of $d$, and the $\varepsilon$-expansion can be used to extrapolate perturbative results to non-perturbative regimes~\cite{Wilson:1971bg,Wilson:1971dh,Wilson:1973jj}. 
However, vectors in non-integer $d$ formally require an infinite-dimensional vector space~\cite{Collins:1984xc}, which allows for non-trivial antisymmetrization of any number of Lorentz indices,
resulting in infinitely many states that decouple or vanish in the limit of integral dimensions. The presence of such \emph{evanescent} states is associated to the violation of unitarity~\cite{Hogervorst:2014rta,Hogervorst:2015akt,Jin:2023fbz}, 
and the behavior of theories in the limit as $d$ approaches integer values has recently been studied in, e.g., Refs.~\cite{Cappelli:2018vir,Zan:2026oyb,Delmastro:2026dpw}. 

In this paper, we will consider the analogous problem of treating the number of flavors ($n$) as a continuous variable~\cite{Morris:1997xj,Cardy:1999zp,Liu:2012ca,Cardy:2013rqg,Shimada:2015gda,Binder:2019zqc,Chlebicki:2020pvo,Gorbenko:2020xya,Grans-Samuelsson:2021uor,Sirois:2022vth,Jacobsen:2022nxs,Nekrasov:2023xzm,Cao:2023psi,Caputa:2025ikn,Henriksson:2025kws}, which provides a simpler setting to study the mechanism behind the evanescence of operators as $n$ approaches integer values.
In particular, Ref.~\cite{Cao:2023psi} 
identified non-perturbative constraints on the operator spectrum in theories with a global $O(n)$ symmetry, dictated by degeneracies in the representation theory at integer $n$. 
The authors showed that evanescent operators can leave the spectrum in two ways. First, an operator may disappear on its own because the dimension of its representation becomes zero; this imposes no constraints. Second, two operators may \emph{annihilate} each other when the character of one irreducible representation (irrep) becomes minus that of the other~\cite{KoikeTerada1987,Cao:2023psi}. In the latter case, the contributions of the two evanescent states to any observable must cancel, which requires their scaling dimensions to be equal. We review the argument of Ref.~\cite{Cao:2023psi} in more detail with explicit examples in Sec.~\ref{sec:2}.

\begin{figure}[h!]
\includegraphics[width=\textwidth, trim=0 0.4cm 0 0, clip]{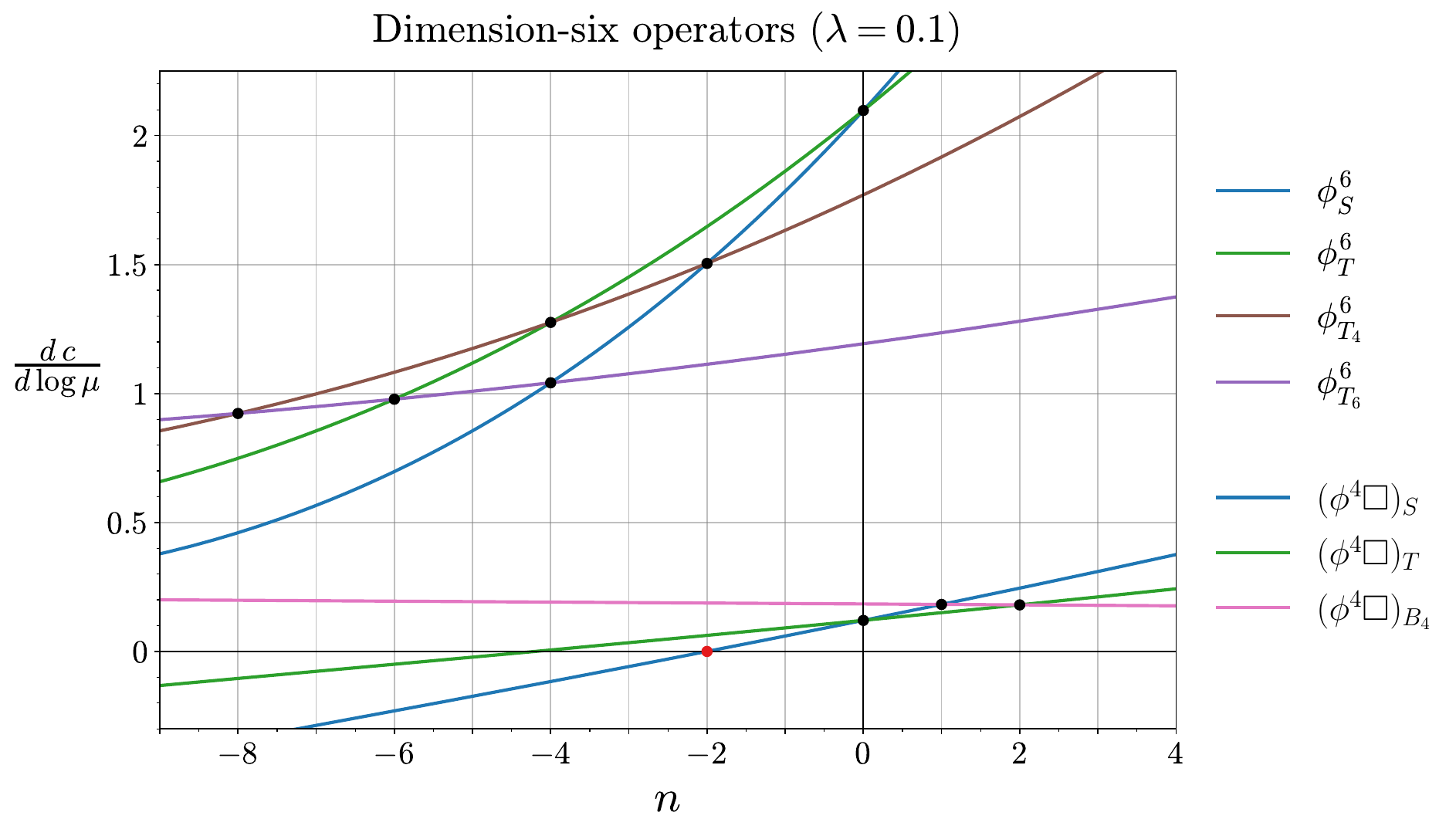}
\hspace{-5mm}
\caption{
Anomalous dimensions of the couplings of the (Lorentz-singlet) dimension-six operators in the $O(n)$-model EFT, defined in Eq.~\eqref{eq:ONLagr}, at five loops as a function of $n$~\cite{Henriksson:2025hwi,Henriksson:2025vyi}.
The top four curves correspond to $\phi^6$-type operators and the lower three curves correspond to operators with field content $\phi^2\partial_\mu\phi\partial^\mu\phi$ (in the free-theory limit).
The subscripts on the operators refer to their $O(n)$ representations, defined in Table~\ref{Tab:multiplicities}.
The marginal $\phi^4$ coupling is chosen to be $\lambda=0.1$ for optimal illustration of the spectrum constraints. 
The anomalous dimensions of the different lines are non-perturbatively constrained to coincide at the black dots (Ref.~\cite{Cao:2023psi} and Sec.~\ref{sec:pertProof}), while the anomalous dimension of the $\phi^4\square$-type singlet operator exactly vanishes at $n=-2$ (red dot, explained in Sec.~\ref{sec:n=-2}).
}
\label{fig:examplesDim6}
\end{figure}

Ref.~\cite{Cao:2023psi} confirmed their spectrum constraints against results in the critical $O(n)$ model (see~\cite{Henriksson:2022rnm} for a review).
A useful framework to test degeneracies in the operator spectrum more extensively is given by the renormalization of general QFTs with arbitrary field content and global-symmetry representations, completed up to four loops in the renormalizable sector in $d=4-2\epsilon$ dimensions~\cite{Mihaila:2013dta,Poole:2019kcm,Steudtner:2020tzo,Steudtner:2021fzs,Bednyakov:2021qxa,Davies:2021mnc,Bednyakov:2021ojn,Steudtner:2024teg}, with recent extensions to effective field theories (EFTs)~\cite{Fonseca:2025zjb,Aebischer:2025zxg,Misiak:2025xzq,Fonseca:2025cls,Guedes:2025sax,Henriksson:2025hwi,Henriksson:2025vyi}.
Using the results from Refs.~\cite{Henriksson:2025hwi,Henriksson:2025vyi}, we plot the anomalous dimensions of the Lorentz-singlet dimension-six operators in all representations in Fig.~\ref{fig:examplesDim6} as a function of $n$.%
    \footnote{In this work, we will work with the $O(n)$ model EFT without tuning the $\phi^4$-coupling $\lambda$ to the conformal fixed-point value, restricting to leading order in the EFT couplings for simplicity. The spectrum constraints are based on representation theory, and are thus valid both in EFT and conformal field theory.}
The spectrum constraints are visible as intersection points between different curves for integer values of $n$.  
While Ref.~\cite{Cao:2023psi} explicitly considered spectrum constraints that occur for positive integers $n$,
Fig.~\ref{fig:examplesDim6} demonstrates that a large number of constraints occur for negative $n$.
One of the main goals of the current work is therefore to explain the origin of these constraints, which we accomplish by relying on the duality between $SO(-n)$ and $Sp(n)$ for even values of $n$~\cite{Penrose:1971,King_1971,MKRTCHYAN1981174,CvitanovicKennedy1982,KoikeTerada1987,Cvitanovic:2008zz,Mkrtchyan:2010tt,Caputa:2013vla,Gurau:2022dbx,Keppler:2023lkb}, as we work out in Sec.~\ref{section3}.

Another observation that Fig.~\ref{fig:examplesDim6} clearly illustrates is that the spectrum constraints relate the anomalous dimensions in intricate patterns. In this way, the theory for generic $n$ is much richer than the theory for specific $n$. This raises the question whether the spectrum constraints can be used to `bootstrap' the scaling dimensions of some operators in terms of others. 
We answer this question affirmatively at leading orders in perturbation theory in Sec.~\ref{Sec:family}, where we reduce the two-loop anomalous dimensions of infinitely many operators to two unknown functions and determine those functions from existing results. 
As our main tool, we rely on the organization of anomalous dimensions in terms of reducible $O(n)$ representations as inspired by the general EFT framework, 
which is well suited to specialization from generic to fixed $n$; 
see e.g.~\cite[Sec.\,5.1]{Henriksson:2025vyi}. 
By scrutinizing the decomposition of reducible representations into irreps in specific examples, we rederive the spectrum constraints in a way that manifests their origin in terms of the involved tensor structures.
Moreover, this perspective directly identifies the pairs of operators that will annihilate each other, which is relevant when there are multiple operators in the same irrep. 
We hope that these insights will prove useful when considering similar spectrum constraints related to the continuation in the number of spacetime dimensions~\cite{Zan:2026oyb}.

Finally, although the spectrum constraints relate anomalous dimensions in different irreps,
we expect that they remain relevant in theories in which one typically restricts to singlet operators, such as the Standard Model EFT.
For example, in such cases it can still be useful to organize operators into irreps of weakly broken symmetries to reveal an approximate block-diagonal mixing structure~\cite{Machado:2022ozb,Renner:2025cmd}. Similar all-loop non-renormalization results can also be obtained by decomposing operators into irreps of accidental symmetries that are broken only at higher orders in an EFT expansion~\cite{Henriksson:2025vyi}.
Even when restricting to singlet operators, we show in Sec.~\ref{sec:3} that evanescence (as $n$ approaches an integer) leaves an imprint on the renormalization-group mixing matrices. More specifically, we use evanescence to predict overall factors of $(n\pm a)$ in the mixing matrices of judiciously chosen operator bases
and we derive a new mechanism for non-renormalization results, both of which we expect to extend beyond the $O(n)$ model. 

After discussing the spectrum constraints in the main text, we conclude in Sec.~\ref{sec:concl} with a brief summary and an outlook on extending the analysis to the operator product expansion and to constraints arising from evanescence under continuation in the number of spacetime dimensions.

\section{\texorpdfstring{\boldmath Spectrum constraints in the $O(n)$ model}{Spectrum constraints in the O(n) model}}\label{sec:2}

In this section, we aim to review the spectrum constraints in the $O(n)$ model that were proposed in Ref.~\cite{Cao:2023psi}. 
For this reason, we first provide a brief review of 
some necessary aspects in the representation theory of the orthogonal group $O(n)$ and 
the symmetric group $S_m$ in Sec.~\ref{sec:2.1RepTheory}, focusing on the construction of operators in reducible and irreducible representations of $O(n)$.
Throughout Sec.~\ref{sec:2.1RepTheory} we assume $n$ to be generic, or equivalently to be an integer that is large enough to avoid low-rank exceptions in decompositions of reducible into irreducible representations. 
Such degeneracies for small values of $n$ are at the heart of the spectrum constraints, which we introduce through a simple example from a novel perspective in Sec.~\ref{sec:inv}, after which we review the general argument of Ref.~\cite{Cao:2023psi} in Sec.~\ref{sec:review}.

The treatment of the $O(n)$ model for generic $n$ 
has previously been discussed in the literature. In particular, we find the discussions in Refs.~\cite{Grans-Samuelsson:2021uor,Cao:2023psi,Henriksson:2025kws} useful. A more mathematical perspective has been provided in Refs.~\cite{Deligne,Binder:2019zqc}. 
We refer to, for example, Ref.~\cite{FultonHarris} for more details on the representation theory in general.

  \subsection
  [Operators in representations of
   \texorpdfstring{$O(n)$ and $S_m$}{O(n) and S\_m}
   ]
  {\boldmath
   Operators in representations of $O(n)$ and $S_m$}
  \label{sec:2.1RepTheory}

We will study operators in irreducible representations of the orthogonal group $O(n)$, working with the $O(n)$-model Lagrangian,
\begin{equation}\label{eq:ONLagr}
    \mathcal{L} = \frac12 \partial_\mu \phi^a \partial^\mu \phi^a 
    -\frac{16\pi^2\,\lambda}{4!} (\phi^a\phi^a)^2 
    + \sum_{\mathcal R\in\operatorname{Irr}(O(n))}\sum_i\,
        c_{\mathcal{R},i}\,
        \mathcal{O}_{\mathcal{R},i}\,,
\end{equation}
where the flavor index $a$ runs from $1$ to $n$ (repeated indices are summed over) and 
we include all independent operators (labeled by $i$) in all irreducible representations of $O(n)$.
Although we add operators with non-zero Wilson coefficients ($c_{\mathcal{R},i}$) to the Lagrangian in an EFT sense, 
we will work at leading order in these couplings, so that operators in different representations do not mix under the renormalization group.
We may include Lorentz non-singlet operators in the Lagrangian in the same way.

Finite-dimensional irreducible representations of $O(n)$ are labeled by Young diagrams, each corresponding to a tensor with a prescribed pattern of symmetrization along rows and antisymmetrization along columns, as well as the subtraction of traces.
For instance, the field $\phi^a$ transforms in the vector representation ($V$) with diagram $\raisebox{0.4mm}{\scalebox{0.35}{$\ydiagram{1}$}}_{\,O(n)}$, while 
$\raisebox{0.4mm}{\scalebox{0.35}{$\ydiagram{2}$}}_{\,O(n)}$ denotes the two-index traceless symmetric representation ($T$) and 
$\,\raisebox{1mm}{\scalebox{0.35}{$\ydiagram{1,1}$}}_{\,O(n)}$ stands for the antisymmetric representation ($A$).%
    \footnote{Aiming for a continuation to general $n$, we will restrict Young tableaux to parity-even representations in this work. 
    Parity-odd representations (for specific $n=\bar n$) can be obtained from analytically continuing representations that antisymmetrize $\bar n$ indices.}
We will call the singlet representation $S$.

Operators in the $O(n)$ model can be built from products of the fields and their derivatives, which generally results in reducible representations. For instance, the product $\phi^a\phi^b$ can be decomposed as
\begin{align}\label{phiphiTensorProduct}
     \phi^a \phi^b = \tfrac{\delta^{ab}}{n} \phi^2
    +\left(\phi^{(a}\phi^{b)} - \tfrac{\delta^{ab}}{n}\phi^2
    \right) + \phi^{[a}\phi^{b]} 
    \quad 
    \leftrightarrow
    \quad 
    V\otimes V \simeq S+ T + A\,,
\end{align}
in terms of the irreducible representations $S$, $T$ and $A$.
In this case, the last operator actually vanishes because the fields commute.
We can therefore restrict to the symmetric tensor product of the vector representation,
\begin{align}\label{eq:sym2Decomp}
&\text{Sym}^2_\otimes(V) \simeq \,
    S+T\,.
\end{align}
For reference, the coefficients in the decomposition of tensor products of $O(n)$ irreps are called Newell--Littlewood numbers~\cite{Newell1951,Littlewood1958,KoikeTerada1987}.

Given that generic operators formed by products of fields and derivatives live in reducible $O(n)$ representations,
it will be advantageous to repackage the $O(n)$-model Lagrangian \eqref{eq:ONLagr} in terms of the representations of the symmetric group $S_m$ for products of $m$ fields and an arbitrary number of derivatives,
\begin{equation}\label{eq:ONLagrGen}
    \mathcal{L} = \frac12 \partial_\mu \phi^a \partial^\mu \phi^a 
    -\frac{16\pi^2\,\lambda^{abcd}}{4!} 
    \phi^a\phi^b\phi^c\phi^d
    + \sum_m\sum_{\mathcal R\in\operatorname{Irr}(S_m)}\sum_i 
        c_{\mathcal{R},i}^{ab\cdots}\,
        \mathcal{O}_{\mathcal{R},i}^{ab\cdots}\,,
\end{equation}
where $\lambda^{abcd}$ is fully symmetric in its indices and the sum over operators includes, for instance, 
$c^{ab}_{\raisebox{0.4mm}{\scalebox{0.25}{$\ydiagram{2}$}}_{\,S_2}}\phi^a \phi^b$, 
which
lives in the trivial representation of the symmetric group $S_2$, with Young tableau
 $\raisebox{0.4mm}{\scalebox{0.35}{$\ydiagram{2}$}}_{\,S_2}$, as dictated by its properties under permutations of the indices. By abuse of notation, we will therefore write the tensor-product decomposition of Eq.~\eqref{eq:sym2Decomp} as
\begin{align}\label{phi2Decomp}
   \phi^a\phi^b \leftrightarrow \,
   \raisebox{0mm}{\scalebox{0.35}{$\ydiagram{2}$}}_{\,S_2} \simeq \,
S+
    \raisebox{0.4mm}{\scalebox{0.35}{$\ydiagram{2}$}}_{\,O(n)} 
   \,.
\end{align}
For operators with derivatives, it is generally more difficult to determine the $S_m$ representations that result in independent operators in the Lagrangian, due to redundancies resulting from field redefinitions and the vanishing of total derivatives in the action.

To illustrate these redundancies, consider the operator with four fields and two contracted derivatives, $\phi^a\phi^b\partial_\mu\phi^c \partial^\mu\phi^d$, which can be decomposed into the $S_4$ representations
$\raisebox{0.4mm}{\scalebox{0.35}{$\ydiagram{4}$}}_{\,S_4}$, 
$\raisebox{1.2mm}{\scalebox{0.35}{$\ydiagram{3,1}$}}_{\,S_4}$ and 
$\,\raisebox{1.2mm}{\scalebox{0.35}{$\ydiagram{2,2}$}}_{\,S_4}$. In this case, only the third representation results in an independent operator \cite{Fonseca:2025zjb} -- see also \cite{Guedes:2025sax} for further discussion on this operator.
This can be understood from the fact that the operators with three symmetrized indices can be related to a total derivative and a term that is redundant under field redefinitions (or at leading order under the equations of motion),
\begin{equation}
    \phi^{(a}
    \phi^b
    \partial_\mu\phi^{c)}\partial^\mu\phi^d
    = \frac13 \partial_\mu\!\left(  \phi^{(a}
    \phi^b\phi^{c)}\partial^\mu\phi^d\right) 
    - \frac13 \phi^{(a}
    \phi^b\phi^{c)}\partial^2\phi^d\,.
\end{equation}
In practice, the determination of the $S_m$ representations that lead to independent Lorentz-singlet operators has been entirely solved through the development of the \texttt{Mathematica} package \texttt{Sim2Int}~\cite{Fonseca:2017lem,Fonseca:2019yya}, which outputs all Young diagrams that are present in a non-redundant Lagrangian.%
\footnote{
After loading \texttt{<<Sym2Int`}~\cite{Fonseca:2017lem,Fonseca:2019yya}, which also imports \texttt{Groupmath}~\cite{Fonseca:2020vke}, the $S_m$ representations of all Lorentz-scalar operators in a minimal basis can efficiently be generated through the simple implementation:
\begin{align*}
&\texttt{gaugeGroup[generalFlavor] \textasciicircum= \{\}; fields[generalFlavor] \textasciicircum= \{\{\textquotedbl$\phi$\textquotedbl, \{\}, \textquotedbl S\textquotedbl, \textquotedbl \,\textquotedbl, n\}\};}\\
&\texttt{GenerateListOfCouplings[generalFlavor, MaxOrder -> 10];}
\end{align*}
here up to ten fields or derivatives for a theory with $n$ scalar fields $\phi^a$, $a=1,\dots,n$.
}

The Lagrangian~\eqref{eq:ONLagrGen} can be viewed as a general EFT Lagrangian for a theory of $n$ scalar fields without flavor symmetry. The operators in this theory without derivatives and up to four fields were renormalized at six loops in Ref.~\cite{Bednyakov:2021ojn}, and a set of operators with derivatives, including non-trivial Lorentz representations, was renormalized up to five loops in Refs.~\cite{Henriksson:2025hwi,Henriksson:2025vyi}.
The $O(n)$ model theory can be obtained from the general EFT by choosing the marginal coupling $\lambda^{abcd} = \frac{\lambda}{3}\left( \delta^{ab}\delta^{cd} + \delta^{ac}\delta^{bd} + \delta^{ad}\delta^{bc}\right)$, while other components of $\lambda^{abcd}$ encode non-trivial irreps following the decomposition
\begin{align}
   \phi^a\phi^b\phi^c\phi^d 
   \quad \leftrightarrow \quad 
   \raisebox{0mm}{\scalebox{0.35}{$\ydiagram{4}$}}_{\,S_4} \simeq \,
S+
    \raisebox{0.4mm}{\scalebox{0.35}{$\ydiagram{2}$}}_{\,O(n)} 
   +
    \raisebox{0.4mm}{\scalebox{0.35}{$\ydiagram{4}$}}_{\,O(n)} 
    \,.
\end{align}
Similarly, the decomposition of the $\phi^4\square$-type operators is
\begin{align}\label{decomposition2,2}
   \phi^a\phi^b\partial_\mu\phi^c
   \partial^\mu\phi^d \quad 
   \leftrightarrow\quad 
    \raisebox{1.2mm}{\scalebox{0.35}{$\ydiagram{2,2}$}}_{\,S_4} \simeq \,
    S 
    +
    \raisebox{0.4mm}{\scalebox{0.35}{$\ydiagram{2}$}}_{\,O(n)} 
    + 
    \raisebox{1.2mm}{\scalebox{0.35}{$\ydiagram{2,2}$}}_{\,O(n)} 
    \,.
\end{align}
In general, such decompositions are given by the Littlewood restriction rule~\cite{10.1098/rsta.1944.0003,Littlewood1950} for the decomposition of a $GL(n)$ irrep into $O(n)$ irreps, using the Schur–Weyl duality to relate the $S_m$ representation to a $GL(n)$ representation. 
Knowledge of the spectrum of operators in the general EFT~\eqref{eq:ONLagrGen} therefore directly determines the spectrum of the $O(n)$ model~\eqref{eq:ONLagr} for generic (or large enough) $n$.

Finally, we note that the dimensions of the reducible representations can be obtained using the hook content formula on the corresponding $GL(n)$ Young tableau, while the dimensions of the $O(n)$ irreps can be determined using the orthogonal hook formula~\cite{ElSamra_1979}. For convenience, we list in Table~\ref{Tab:multiplicities} the dimensions for the $O(n)$ irreps that we will discuss in this work.\footnote{We note that the term dimension is used both for the scaling dimension of operators and for the dimension of representations. The intended meaning should be clear from context.}

\subsection{Invitation to the spectrum constraints}\label{sec:inv}

The decomposition of 
reducible $O(n)$ representations for generic $n$, considered in the previous subsection, is subject to low rank exceptions when particular $O(n)$ irreps in the decomposition cease to exist. 
This leads to constraints on the spectrum of the $O(n)$ model when the anomalous dimensions are viewed as a function of $n$. 
We will first illustrate this in a simple example, before discussing the argument of Ref.~\cite{Cao:2023psi} in the next subsection.

The decomposition of the operator (suppressing the subscript $\raisebox{0.4mm}{\scalebox{0.35}{$\ydiagram{2}$}}_{\,S_2}$ on the coupling)
\begin{align}\label{eq:phi^2}
    \mathcal{O}_{\phi^2} = c^{ab}\phi^a \phi^b
\end{align}
into the $O(n)$ irreps $S$ and $T$ (see Eq.~\eqref{phi2Decomp}) requires $n$ to be large enough to accommodate both representations. In this case, the $T$ representation requires $n\geq 2$. 
More generally, the sum of lengths of the first two columns of a Young tableau cannot exceed $n$.
For $n=1$, the $T$ representation vanishes, as can be observed from its dimension, $d_T = \tfrac12(n-1)(n+2)$,
going to zero. 
Since $d_T\big|_{n=0}=-1$ (and $d_S=1$ for all $n$), the total number of components of the symmetric matrix,
\begin{equation}
    \tfrac12n(n+1) = d_S + d_T\,,
\end{equation}
goes to zero for $n=0$.
This hints at a cancellation between the two irreps.

\begin{table}[t]
\centering
\begin{tabular}{ccc}
\hline
Representation ($\mathcal{R}$) & Young tableau & Dimension ($d_\mathcal{R}$) \\
\hline\\[-4mm]
$S$ & $\emptyset$ & $1$ \\[6pt]
$V$ & $\raisebox{0mm}{\scalebox{0.35}{$\ydiagram{1}$}}$ & $n$ \\[6pt]
$T$ & $\raisebox{0.5mm}{\scalebox{0.35}{$\ydiagram{2}$}}$ & $\dfrac{(n-1)(n+2)}{2}$ \\[10pt]
$T_k$ & 
\hspace{1mm}$\raisebox{1mm}{$\underbrace{\scalebox{0.4}{$
\begin{array}{c}
\begin{array}{|c|}
\hline
\rule{0pt}{2.4ex}\rule{2.4ex}{0pt} \\
\hline
\end{array}
\end{array}
\;\raisebox{-1mm}{\scalebox{1.8}{$\cdots$}}\,
\begin{array}{c}
\begin{array}{|c|}
\hline
\rule{0pt}{2.4ex}\rule{2.4ex}{0pt} \\
\hline
\end{array}
\end{array}
$}\rule[-3pt]{0pt}{0pt}}_{k}$}$
&
$\dbinom{n+k-1}{k} - \dbinom{n+k-3}{k-2}$
\\[6mm]
$A$ & $\raisebox{1mm}{\scalebox{0.35}{$\ydiagram{1,1}$}}\,$ & $\dfrac{n(n-1)}{2}$ \\[10pt]
$A_k$ & 
$\hspace{3.7mm}\left.
\raisebox{0.5mm}{\scalebox{0.4}{$
    \begin{array}{c}
    \begin{array}{|c|}
    \hline
    \rule{0pt}{2.4ex}\rule{2.4ex}{0pt} \\
    \hline
    \end{array} \\[-2mm]
    \scalebox{1.8}{$\vdots$} \\[-0ex]
    \begin{array}{|c|}
    \hline
    \rule{0pt}{2.4ex}\rule{2.4ex}{0pt} \\
    \hline
    \end{array}
    \end{array}
$}}
\right\} \scriptstyle{k}$
& $\dbinom{n}{k}$ \\[4mm]
$B_4$ & $\raisebox{1mm}{\scalebox{0.35}{$\ydiagram{2,2}$}}$ & $\dfrac{n(n+1)(n+2)(n-3)}{12}$ \\[6pt]
\hline
\end{tabular}
\caption{Irreducible representations of $O(n)$ with their Young tableaux and dimensions.}
\label{Tab:multiplicities}
\end{table}

Such  a cancellation becomes manifest upon inspecting the explicit tensor structures of the irreps, $c^{ab} = c_S^{ab} + c_T^{ab}$, with 
\begin{align}
    c_S^{ab} &= \delta^{ab} \textstyle\sum_x c^{xx}\,,
    &c_T^{ab} &= c^{ab} - \tfrac1n c_S^{ab}\,,
    \label{eq:cT}
\end{align}
The appearance of $c_S$ on the right-hand side of $c_T$ in \eqref{eq:cT} corresponds to the subtraction of the trace. 
Other representations of $O(n)$ are also interconnected through the subtraction of traces, which is the underlying mechanism for the spectrum constraints.
In the $O(n)$ model, the anomalous dimension of $c^{ab}$ can be computed before the decomposition into irreducible representations.
Since $\delta^{ab}$ is the only invariant tensor in the $O(n)$ model, the anomalous dimension of $c^{ab}$ takes the general form
\begin{align}\label{gen}
    \frac{d\,c^{ab}}{d\log\mu}&=f_1(\lambda,n) \,c^{ab}
    + f_2(\lambda,n)\,\delta^{ab}\textstyle\sum_x c^{xx}\,,
\end{align}
where $f_1$ and $f_2$ are unknown functions of the coupling $\lambda$ and the number of fields $n$. These can be computed, for example, in perturbation theory, in which case $f_1$ and $f_2$ are polynomial in $\lambda$ and $n$ at each order. From \eqref{gen}, we can extract the anomalous dimensions of $c_S^{ab}$ and $c_T^{ab}$ in terms of $f_1$ and $f_2$ by projecting to the specific irreps,
\begin{align}\label{2.5}
    \frac{d\,c_S^{ab}}{d\log\mu}&=\big(
    f_1(\lambda,n) 
    +n\,f_2(\lambda,n)\big)\,
    c_S^{ab}\,,
    &
    \frac{d\,c_T^{ab}}{d\log\mu}&=
    f_1(\lambda,n) 
    \,
    c_T^{ab}\,,
\end{align}
which makes manifest that these anomalous dimensions agree for $n=0$\,,
\begin{align}\label{eq:constrph2}
    \Delta_S \Big|_{n=0} = \Delta_{T}\Big|_{n=0}\,,
\end{align}
where $\Delta_\mathcal{O}$ refers to the full (i.e.\ classical plus anomalous) dimension of the corresponding operator.
Although we focused on operators with two fields here, our derivation of this spectrum constraint can straightforwardly be generalized to arbitrary operators in the same way.
The same degeneracy was previously argued to be necessary to ensure that the partition function is unity in the limit $n=0$~\cite{Cardy:1999zp} (see also~\cite{Shimada:2015gda}). We discuss the physical meaning of the $O(n)$ model for $n=0$ in Sec.~\ref{section3}~\cite{deGennes:1972zz}.

In the above derivation, we relied on the fact that $f_2$ is regular as $n\to0$, which is guaranteed at each finite order in perturbation theory, or if $dc^{ab}/d\log\mu$ is regular as $n\to0$.
The argument of Ref.~\cite{Cao:2023psi} derives the same spectrum constraints non-perturbatively, which we will now review.

\subsection[Spectrum constraints in the \texorpdfstring{$O(n)$}{O(n)} model]{\boldmath
   Spectrum constraints in the $O(n)$ model}
\label{sec:review}

In the previous section, we observed a spectrum constraint as a consequence of the negative dimensionality of a representation, ${d_T=-1}$ at $n=0$, when we write $d_T$ as a function of $n$.
A compact way to capture more representation-theoretic information is through characters, which can also be continued to general $n$. 
The character of an $O(n)$ representation $\R$, $\chi_\R^\supON(x_i)$, is a function of 
bookkeeping variables $x_i$ that encode the eigenvalues of a group element, with $i=1,...,r$ up to the rank $r=\lfloor\tfrac{n}{2}\rfloor$\,.%
    \footnote{
    Strictly speaking, we restrict the $O(n)$ characters to $SO(n)$; this is sufficient because we consider only
parity-even tensor representations constructed without the Levi–Civita tensor, while full $O(n)$ characters would additionally distinguish their transformation under reflections.
    For example, for a group element $g\in SO(2)\subset O(2)$, we have $x=e^{i\theta}$, where $\theta$ is the rotation angle. In this case, \begin{equation*}
    \chi_V^{O(2)}(x)=\text{Tr}\begin{pmatrix}
        \cos\theta & -\sin\theta \\
        \sin\theta & \cos\theta 
        \end{pmatrix}=2\cos\theta=x+x^{-1}\,.
        \end{equation*}
    }
In particular, the singlet and vector representations have characters
\begin{align}
&\chi_S^\supON(x_i) = 1 \,,
&& \text{and}
&& \chi(x_i) \equiv \chi_V^\supON(x_i) = \frac{1-(-1)^n}{2}+\sum_{j=1}^{r} \left(x_j + \tfrac1{x_j}\right) ,
\end{align}
where for brevity we will suppress 
the subscript ${}_V$ whenever we refer to $\chi_V^\supON$.
The characters of other representations can be written in terms of the character of the vector representation~\cite{KoikeTerada1987} (see also \cite{Cao:2023psi}); for example
\begin{align}
    \chi_{T}^\supON(x_i) &=
    \tfrac12 [\chi(x_i)]^2+\tfrac12 \chi(x_i^2) 
    -\chi_S^\supON(x_i)\,,\nn\\
    \chi_{A}^\supON(x_i) &=
    \tfrac12 [\chi(x_i)]^2-\tfrac12 \chi(x_i^2) 
    \,,\nn\\
    \chi_{B_4}(x_i) &= 
    \tfrac1{12} [\chi(x_i)]^4
    +\tfrac14 [\chi(x_i^2)]^2
    -\tfrac13 \chi(x_i)\chi(x_i^3)
    -\chi_{T}(x_i)
    -\chi_S(x_i)\,,
    \label{eq:charGenn}
\end{align}
where $x_i^k$ is shorthand for 
$\{x_1^k,...,x_r^k\}$ and $B_4$ refers to the representation with Young tableau $\,
\raisebox{1mm}{\scalebox{0.35}{$\ydiagram{2,2}$}}_{\,O(n)}$.
Tensor products of representations are efficiently encoded in the characters, such as 
\begin{equation}
[\chi(x_i)]^2 = \chi_S + \chi_T(x_i) + \chi_A(x_i)\,,
\end{equation}
cf.~\eqref{phiphiTensorProduct}.
Moreover, the dimension of a representation can be obtained from its character evaluated at the identity, using $\chi(1)=n$. 

Let us consider the following partition function weighted by the character of the representations, following~\cite{Cao:2023psi},
\begin{align}\label{eq:Z}
    {Z}_{O(n)}(q,x) &=
    \sum_{\mathcal R\in\operatorname{Irr}(O(n))}
    {\chi_{}}_{\mathcal{R}}(x) 
    \,
    \sum_i q^{\Delta_i}
    \,,
\end{align}
where we sum over all $O(n)$ irreps $\R$ and all states in each representation, labeled by~$i$, and we treat $q$ as a bookkeeping variable. The exponent $\Delta_i$ is the scaling dimension of the state $i$, i.e.\ the classical plus anomalous dimension.
For a particular integer $n=\bar n$ of interest, the 
partition function can be evaluated in two different ways. Either we work with the $O(\bar n)$ model throughout and sum over only those representations and states that exist for $n=\bar n$. Alternatively, we first consider the $O(n)$ model for general $n$ and take the limit $n\to\bar n$ at the end. 
Assuming continuity in $n$, these two approaches should agree for any observable that can be computed in both ways (e.g.~\cite[Sec.\,2.5]{Henriksson:2025kws}). 

Given that the character of the vector representation exists for any $n\geq 1$, the characters of other irreps can similarly be continued to any $n$ when written in terms of $\chi_V(x_i)$, as in \eqref{eq:charGenn}. However, for specific values of $\bar n$, these characters may \emph{specialize} to zero, or to plus or minus the character of another irrep~\cite{KoikeTerada1987}. These specialization rules were worked out in detail in~\cite{Cao:2023psi}. When the character of a representation specializes to zero for $n=\bar n$, the contribution to the partition function~\eqref{eq:Z} evaluated at $n=\bar n$ (without the representation in question) and the limit $n\to \bar n$ naturally agree. 
In this case, the evanescent states individually leave the spectrum, without any constraints on their scaling dimensions.
On the other hand, when a character specializes to minus the character of another irrep, spectrum continuity necessitates that two associated states \emph{annihilate} each other~\cite{Cao:2023psi}. Said otherwise, in this case the evanescent states leave the spectrum in pairs, such that their contribution to the partition function~\eqref{eq:Z} cancels. This is the origin of the spectrum constraints in the $O(n)$ model~\cite{Cao:2023psi}.

\paragraph{Example at dimension six.}
Let us illustrate the specialization rules and spectrum constraints for the dimension-six Lorentz-scalar operator with four fields and two derivatives,%
    \footnote{For concreteness, we rely here on an explicit expression of the operator in terms of the fields, assuming a perturbative definition of the theory. This is not necessary for the argument of~\cite{Cao:2023psi}, which only relies more abstractly on the states in specific irreps. At dimension six, there are also operators with field content $\phi^a\phi^b\phi^c\phi^d\phi^e\phi^f$; we discuss the effects of operator mixing in more detail in Sec.~\ref{sec:pertProof} and Sec.~\ref{sec:3}.}
\begin{align}\label{opeq2.1}
    \mathcal{O}_{\phi^4\square} = c^{abcd} \phi^a \phi^b \partial_\mu \phi^c \partial^\mu \phi^d\,.
\end{align}
We showed in~\eqref{decomposition2,2} that the primary operators live in the $S_4$ representation with Young diagram \,$\raisebox{1.2mm}{\scalebox{0.35}{$\ydiagram{2,2}$}}_{\,S_4}$, which decomposes into the 
$O(n)$ irreps $S$, $T$ and $B_4$ 
for generic $n$ (specifically $n\geq 4$). 
This reflects the fact that the total number of parameters in $c^{abcd}$ is given by 
\begin{align}\label{Dimphi4box}
    \frac{n^2}{12}(n+1)(n-1)
    = d_S + d_{T} + d_{B_4}
    \,,
\end{align}
which we take to hold for any $n$ (see Table~\ref{Tab:multiplicities} for $d_\mathcal{R}$).

For integers below the stable range $n\geq 4$, the operators disappear from the spectrum in stages until all operators are evanescent for $n=1$, for which the total number of physical parameters~\eqref{Dimphi4box} is zero.
For $n=3$, we have that $\chi^{O(3)}_{B_4}(x_i)=0$ and hence the operator in the $B_4$ representation is evanescent by itself. For $n=2$, the characters in Eq.~\eqref{eq:charGenn} specialize to~\cite{KoikeTerada1987,Cao:2023psi}
\begin{align}
    \chi_{B_4}^{O(2)} = -\chi_T^{O(2)}\,,
\end{align}
leading to the spectrum constraint~\cite{Cao:2023psi}
\begin{align}\label{eq:constr2}
    \Delta_T \Big|_{n=2} = \Delta_{B_4}\Big|_{n=2}\,.
\end{align}
In this case, both the operator in the $B_4$ representation and that in the $T$ representation are evanescent, even though the $T$ representation exists for $n=2$. They leave the spectrum in pairs due to the annihilation mechanism.

Similarly, for $n=1$, we have~\cite{Cao:2023psi} 
\begin{align}\label{eq:constr4}
     \chi_T^{O(1)} = 0\,,
    \quad \chi_{B_4}^{O(1)} = -\chi_S^{O(1)}
    \quad\implies \quad 
        \Delta_S \Big|_{n=1} = \Delta_{B_4}\Big|_{n=1}\,,
\end{align}
where the theory with $O(1)$ symmetry is the $Z_2$-symmetric single-scalar theory.
In this case, all states are evanescent but the singlet and the $B_4$ representation annihilate each other, leading to a constraint on the corresponding scaling dimensions. It should be emphasized that the $O(n)$ singlet representation exists for $n=1$. However, the operator in the $S_4$ representation 
$\raisebox{1.2mm}{\scalebox{0.35}{$\ydiagram{2,2}$}}_{\,S_4}$
vanishes for $n=1$, and it can thus not accommodate a singlet component.

Finally, even though the vector representation exists only for integers $n\geq1$, it is natural to continue to characters to $n=0$ by setting $\chi^{O(n=0)}_V=0$, thereby extending the spectrum constraints of Ref.~\cite{Cao:2023psi} to $n=0$, noting that the $O(n)$ model describes self-avoiding random walks in the limit $n\to0$~\cite{deGennes:1972zz}.
This leads to
\begin{align}
    &\chi_{B_4}^{O(n=0)} = 0\,,
    && \chi_T^{O(n=0)} = -\chi_S^{O(n=0)}
    \qquad \implies \qquad
    \Delta_S \Big|_{n=0} = \Delta_{T}\Big|_{n=0}\,,
    \label{eq:constr3}
\end{align}
which is the spectrum constraint that we identified in Sec.~\ref{sec:inv}, albeit for a different state. The generalization of the spectrum constraints to $n=0$ and negative $n$ (to be derived below) are important outcomes of the present work.

\paragraph{Explicit results.}
The above spectrum constraints 
can be explicitly confirmed against existing results in the literature. 
At mass dimension six, besides the $\phi^4\square$-type operators, there exist operators with six fields (schematically: $\phi^6$), which include the singlet and $T$ representations, as well as the four- and six-index traceless symmetric irreps, $T_4$ and $T_6$, respectively. 
Their anomalous dimension mixing matrices have been computed up to five loops in
Refs.~\cite{Henriksson:2025hwi,Henriksson:2025vyi} (see also \cite{RoosmaleNepveu:2024zlz} for results in the singlet representation),{\allowdisplaybreaks
\begin{align}
    \frac{d}{d\log\mu} \pmatrixx{c_{\phi^6}^{(S)}\\c_{\phi^4\square}^{(S)}} &= 
    \pmatrixx{(14+n)\lambda+... 
    &
    (n-1)\left[-\tfrac{10}{9}\lambda^3+...\right]
    \\
    (n+2)(n+4)\left[\tfrac{5}{81}\lambda^3 + ...\right]
    &
    (n+2)\left[\tfrac{2}{3}\lambda+...\right]}
    \pmatrixx{c_{\phi^6}^{(S)}\\c_{\phi^4\square}^{(S)}}
    \,,
    \nn\\[2mm] 
    \frac{d}{d\log\mu} \pmatrixx{c_{\phi^6}^{(T)}\\c_{\phi^4\square}^{(T)}} &= 
    \pmatrixx{\tfrac23(21+n)\lambda+... 
    &
    (n-2)\left[-\tfrac{5}{27}\lambda^3+...\right]
    \\
    (n+4)(n+6)\left[\tfrac{5}{81}\lambda^3 + ...\right]
    &
    \tfrac{1}{3}(n+4)\lambda+...}
    \pmatrixx{c_{\phi^6}^{(T)}\\c_{\phi^4\square}^{(T)}}
    \,,
    \nn\\[2mm]
    \frac{d\, c^{(B_4)}_{\phi^4\square}}{d\log\mu} &= \left(2\lambda+...\right)c^{(B_4)}_{\phi^4\square}\,,
    \nn\\[2mm]
    \frac{d \,c^{(T_4)}_{\phi^6}}{d\log\mu} &= \left(\tfrac13(n+38)\lambda+...\right)c^{(T_4)}_{\phi^6}\,,
    \nn\\[2mm]
    \frac{d \, c^{(T_6)}_{\phi^6}}{d\log\mu} &= \left(10\lambda+...\right)c^{(T_6)}_{\phi^6}\,,
    \label{2.17}
\end{align}
where} we only show the leading contribution in each entry. 
To visualize the spectrum constraints, we have plotted the five-loop anomalous dimensions
in Fig.~\ref{fig:examplesDim6} as a function of $n$ for $\lambda=0.1$.%
    \footnote{The chosen value $\lambda=0.1$ is relatively large, which means that the perturbative expansion is not reliable. For instance, the anomalous dimension of one of the singlet operators numerically evaluates to
    $$
        \gamma_{\phi^6}^{(S)}=
        14\, \lambda
        -63\, \lambda ^2
        +689.109\, \lambda ^3
        -9769.61\, \lambda ^4
        +161508\, \lambda ^5\,,
    $$
    for $n=0$. 
    Nevertheless, Fig.~\ref{fig:examplesDim6} serves to visualize the spectrum constraints, because they are satisfied order by order in perturbation theory, as long as the $n$-dependence is kept exact at each order.
    }
For the $S$ and $T$ irreps, we plotted the eigenvalues of the mixing matrices.

In general, the spectrum constraints of Eqs.~(\ref{eq:constr2}, \ref{eq:constr4}, \ref{eq:constr3}), constrain the eigenvalues of the mixing matrices. 
However, we remark that the mixing problem simplifies for the relevant values of $n$: for $n=1$, one of the off-diagonal entries in the mixing matrix of singlet operators vanishes, such that the (2,2) entry directly agrees with $d c_{\phi^4\square}^{(B_4)}/d\log\mu$. (There is no corresponding $\phi^6$-type operator in the $B_4$ representation.)
The same mechanism occurs for $n=2$ in the mixing matrix of the operators in the traceless symmetric representation. 
For $n=0$, both eigenvalues of the mixing matrices of the operators in the $S$ and $T$ irreps agree, while the matrices do not become exactly equal. This is due to an arbitrary relative normalization of the operators, which could be chosen such that the matrices do agree.

In Sec.~\ref{sec:3} we analyze the patterns in the mixing matrices due to evanescence in more generality, predicting the overall factors of 
$(n-a)$ (with $a\in\mathbb{N}$), as indicated by square brackets in \eqref{2.17}.
In addition, the mixing matrices contain various factors of $(n+a)$  which relate to spectrum constraints at negative $n$, as can be observed from the many intersection points in Fig.~\ref{fig:examplesDim6} in this region. 
We discuss these spectrum constraints in the next section, while the factors of $(n+2)$ that lead to the vanishing of the anomalous dimension of the $(\phi^4\square)_S$ operator at $n=-2$ (red dot in Fig.~\ref{fig:examplesDim6}) will be discussed in Sec.~\ref{sec:n=-2}.

\section{\texorpdfstring{\boldmath Spectrum constraints for negative $n$}{Spectrum constraints for negative n}}
\label{section3}

We ended the previous section by observing additional patterns in the explicit mixing matrices and new spectrum constraints for negative $n$, beyond the predictions covered by Ref.~\cite{Cao:2023psi}. One of the main goals of this work is to explain and leverage these additional constraints.
In this section, we will show how degeneracies in the spectrum for negative even values of $n$ can be explained using the existing duality between the $O(n)$ and the $Sp(-n)$ groups.%
    \footnote{While the negative-dimension duality is usually stated as a correspondence between $SO(n)$ and $Sp(-n)$, we restrict to parity-even operators (without Levi-Civita tensors). In this case we treat the duality as relating $O(n)$ and $Sp(-n)$.
}
The duality originates from the observation~\cite{King_1971,CvitanovicKennedy1982} that for any fully-contracted $O(n)$-invariant quantity there exists a corresponding $Sp(n)$-invariant counterpart, obtained by exchanging the symmetrizations and antisymmetrizations. Their values are then related by the substitution $n\to -n$.
To further motivate our interest in this duality, we briefly introduce some physical examples of $Sp(n)$-invariant theories of anticommuting scalar fields, an instance of which will play a key role in the following sections. 

Prominent examples of phenomena modeled through $O(n)$ invariant scalar QFTs with negative $n$ values come from statistical mechanics. Close to the so-called depinning transitions, the behavior of charge density waves in disordered solids is related to the $O(n)$ symmetric $\phi^4$ model in the limit $n\rightarrow -2$ \cite{Wiese_2019}. The same $n\rightarrow -2$ limit can be used to describe loop-erased random walks; random walks in which loops are erased as soon as they are formed \cite{Wiese:2018dow,Wiese_2019}. Models with symplectic fermions were considered as belonging to a class of conformal field theories whose correlation functions have logarithmic branch cuts \cite{Kausch_2000}, believed to be important for the theory of critical polymers and percolation.

Another application of Euclidean $Sp(n)$ vector models with anticommuting scalars is to holographic realizations of the dS/CFT correspondence~\cite{Anninos:2011ui,Fei:2015kta}. In this less common version of holography, a higher-spin theory of gravity in $(d+1)$-dimensional de Sitter space (dS) is dual to a non-unitary $d$-dimensional conformal field theory living on the spacelike boundary of dS at future timelike infinity. While these physical applications are not the primary focus of our work, it would be interesting to see how the spectrum constraints apply in these cases.

The limit $n\rightarrow 0$  also has a physical meaning, as it can model self-avoiding random walks and polymers \cite{deGennes:1972zz,Cardy:1996xt,Pelissetto:2000ek}. We mention this limit as some of the degeneracies we describe appear precisely for this value of $n$ . However, while we can construct a Lagrangian for the $O(-n)$ theory by using $Sp(n)$ symmetric anticommuting scalar fields, we cannot write a Lagrangian using zero fields. A possible reformulation of the $n\rightarrow 0$ model can be realized by an equal number of commuting and anticommuting fields following Ref.~\cite{McKane:1979rm}. 
We leave a deeper analysis of models with $n$ commuting and $m$ anticommuting fields for future work.

In Sec.~\ref{sec:duality}, we review the duality between $O(-n)$ and $Sp(n)$ for even integers $n=2N$, focusing on the relation between the Lagrangians with these global symmetries. In Sec.~\ref{sec:constraintsSp}, we then analyze the spectrum constraints that occur in the $O(2N)$ model at negative $N$, followed by 
an alternative derivation of the spectrum constraints in Sec.~\ref{sec:pertProof}. This derivation applies at all orders in perturbation theory and it provides additional insight into perturbative computations.

\subsection
  [The duality between \texorpdfstring{$O(-n)$}{O(-n)} and \texorpdfstring{$Sp(n)$}{Sp(n)}]
  {\boldmath The duality between $O(-n)$ and $Sp(n)$}
\label{sec:duality}

The anomalous dimensions in renormalizable $Sp(2N)$ models involving anticommuting scalars can be calculated from the results of $O(2N)$ models by replacing $N \rightarrow -N$~\cite{Fei:2015kta,Fraser-Taliente:2026iuj}. An early example of this appeared in \cite{Parisi:1979ka}, where Parisi and Sourlas suggested that a Grassmann vector space of dimension $n$ can be interpreted as an ordinary vector space of dimension $-n$. Following similar arguments, we will argue that this duality can be extended to the anomalous dimensions of operators with arbitrary mass dimension.%
    \footnote{Using anticommuting fields in \texttt{qgraf}~\cite{Nogueira:1991ex} and the $R^*$ algorithm described in \cite{Henriksson:2025hwi}, we explicitly computed the anomalous dimensions of Lorentz-singlet operators at dimension six up to four loops in the $Sp(n)$ model. We confirmed that the results equal those in the $O(n)$ model upon substituting $n$ by $-n$, as follows from the argument presented in this section.}
We will show that the coincident anomalous dimensions at certain negative values of $N$ in the $O(2N)$ model originate from degeneracies in the representation theory of $Sp(2N)$.
 This extends the argument of Ref.~\cite{Cao:2023psi} in the $O(n)$ model to negative values of $n$.

These observations rely on the existence of a map whose image for a pair of operators in the $O(2N)$ and $Sp(2N)$ theories is the same operator in a complex scalar theory with a global $U(N)$ symmetry and commuting and anticommuting scalar fields, respectively. As a consequence, any observable that can be computed perturbatively from Feynman diagrams will have the same expression in the two theories, up to the substitution $n\rightarrow -n$. In what follows, we show how this mechanism works in concrete examples.

\paragraph{\boldmath Mapping renormalizable $O(-2N)$ to $Sp(2N)$ theories in $d=4$.}
The renormalizable parts of the $O(2N)$ and $Sp(2N)$ model Lagrangians are, respectively, 
\begin{align}\label{eq:Ren_Lagr}
    \mathcal{L}_{O(2N)} &= 
    \frac12 \delta^{ab}\partial_\mu \phi^a \partial^\mu \phi^b 
    -\frac{16\pi^2\,\lambda}{4!} (\delta^{ab}\phi^a\phi^b)^2\,,
    \\
    \mathcal{L}_{Sp(2N)} &= 
    \frac12 J_{ab} \partial_\mu  \varphi^a \partial^\mu \varphi^b 
    -\frac{16\pi^2\,\lambda}{4!} (J_{ab}\varphi^a\varphi^b)^2\,,\label{eq:Ren_Lagr_Sp}
\end{align}
where we keep only operators that are singlet under $O(2N)$ or $Sp(2N)$ transformations and we take the massless limit for simplicity.
The flavor index contraction in the latter theory is performed by the symplectic matrix,
\begin{equation}
J_{ab} = 
\begin{pmatrix}
0_{N\times N}&- i\, \mathbf{1}_{N\times N}\\
i \, \mathbf{1}_{N\times N}&0_{N\times N}
\end{pmatrix}_{ab}\,,
\label{eq:symp_J}
\end{equation}
which differs only by an overall phase from the canonical real symplectic form and defines the same group of transformations preserving the antisymmetric bilinear form. Alternatively, one could redefine $\varphi^a\to i\varphi^a$ for $a\in \{1,\dots,N\}$ to obtain a real kinetic term.
Since $J_{ab}$ is antisymmetric, the fields $\varphi$ are anticommuting.

As written in Eqs.~\eqref{eq:Ren_Lagr} and \eqref{eq:Ren_Lagr_Sp}, the relation between the $O(2N)$ and $Sp(2N)$ models is not manifest, since contractions are performed with different invariant tensors. 
To expose this duality, we adapt the method of \cite{Fei:2015kta} by considering a redefinition from real to complex scalar fields. For example, the commuting scalars can be rewritten as
\begin{align}\label{eq:Complex_Map}
\phi^k &= \frac{\Phi^k+\bar{\Phi}^k}{\sqrt{2}}\,,
&
\phi^{k+N} &= \frac{\Phi^k-\bar{\Phi}^k}{\sqrt{2}i}\,,\\[3mm]
    \Phi^k &= \frac{\phi^k + i \phi^{k+N}}{\sqrt{2}}\,,
    & \bar \Phi^k &= 
    \frac{\phi^k - i \phi^{k+N}}{\sqrt{2}}\,,
    \label{eq:InverseMap}
\end{align}
for $k \in \{1,\dots,N\}$. The anticommuting scalars in the $Sp(2N)$ model can be mapped to the complex basis in the same way.%
    \footnote{In this paper we consider a construction that is slightly different from the one in \cite{Fei:2015kta}. We chose our map such that it gives the same expression for single fields $v_a \phi^a$ and $v_a\varphi^a$ in both cases.}
In both cases, Eq.~\eqref{eq:Ren_Lagr} then becomes 
\begin{equation}\label{eq:Ren_Lagr_C}
    \mathcal{L} = \partial_\mu \Phi^\alpha \partial^\mu \bar{\Phi}^\alpha 
    -\frac{16\pi^2\,\lambda}{3!} (\Phi^\alpha \bar{\Phi}^\alpha)^2\,,
\end{equation}
where $\alpha \in \left\lbrace 1,\ldots,N\right\rbrace$. 
Flavor contractions are now performed using delta functions in both theories, but the ordering of the fields is important for the sign in the anticommuting case.
To show the correspondence between the $O(-2N)$ model and the $Sp(2N)$ model, we thus need to argue that the $U(N)$ theory of Eq.~\eqref{eq:Ren_Lagr_C} is related to itself upon exchanging symmetrizations and antisymmetrizations and substituting $N\to-N$, which corresponds to a known property of $U(N)$, e.g.~\cite{Cvitanovic:2008zz}.

In the theory with commuting fields with Lagrangian~\eqref{eq:Ren_Lagr_C}, the Feynman rule of the $(\Phi\bar\Phi)^2$-vertex is proportional to $\delta^{\alpha\beta}\delta^{\gamma\delta} + 
\delta^{\alpha\delta}\delta^{\gamma\beta}$, while it is proportional to 
$\delta^{\alpha\beta}\delta^{\gamma\delta} - 
\delta^{\alpha\delta}\delta^{\gamma\beta}$ with anticommuting fields. 
To disentangle these minus signs from those arising from closed loops of Grassmann-valued fields, it is useful to introduce an auxiliary commuting field, $\chi$, and rewrite~\eqref{eq:Ren_Lagr_C} as
\begin{align}
    \mathcal{L} = \partial_\mu \Phi^\alpha \partial^\mu \bar{\Phi}^\alpha 
    + \frac12 \chi^2
    + C\,\chi\,\Phi^\alpha \bar \Phi^\alpha\,,
\end{align}
where $C=4\pi\sqrt{\lambda/3}$ is chosen such that Eq.~\eqref{eq:Ren_Lagr_C} is reproduced upon integrating out $\chi$. 
In the resulting theory, the flavor contractions are then aligned with the loops of $\Phi$ fields;
every closed flavor chain produces a trace
\begin{equation}
    \delta^{\alpha_1 \alpha_2}
    \delta^{\alpha_2\alpha_3}
    \cdots
    \delta^{\alpha_\ell \alpha_1}
    =
    \delta^{\alpha_1 \alpha_1}
    =
    N,
\end{equation}
or, equivalently, a factor $n/2$ in the original notation. Importantly, exactly the same loops result in an additional minus sign if the fields are anti-commuting.

For vacuum diagrams, the only source of relative signs arises from the closed flavor traces. However, when there are external fields, additional signs arise from their ordering in Feynman diagrams. 
If $L$ is the number of closed components and $\pi$ is the permutation of external flavor lines induced by the open components, the anticommuting contraction differs from the commuting one by
\begin{equation}
    (-1)^L\operatorname{sgn}(\pi)\,,
\end{equation}
where factor $(-1)^L$ combines with the flavor trace $N^L$ to give $(-N)^L$.
As a result of the factor of $\operatorname{sgn}(\pi)$, the role of symmetrizations and antisymmetrizations of the external indices is exchanged between the commuting and anticommuting theories. 
In this way, 
the Young projector associated with a tableau $\lambda$ is mapped to the projector associated with
the transposed tableau $\lambda^{\mathrm T}$, which is exactly how representations of $O(2N)$ map to those of $Sp(2N)$ under the substitution $N\to-N$~\cite{King_1971,CvitanovicKennedy1982,Mkrtchyan:2010tt}. 
Thus while individual diagrams may contain additional signs, the sum over diagrams obeys the expected correspondence between the $O(-2N)$ model and the $Sp(2N)$ model.

\paragraph{\boldmath Mapping composite operators from $O(-2N)$ to $Sp(2N)$ theories.}

Let us illustrate the application of the duality for other operators. 
The main conclusion from the previous paragraph is that given a representation of $O(2N)$ specified by its Young tableau $\lambda$, we can identify the corresponding representation of $Sp(2N)$, with tableau $\lambda^T$, by exchanging the role of symmetrization and antisymmetrization.
The dimensions of the representations in the two theories are then related via \cite{King_1971,Cvitanovic:2008zz}
\begin{equation}
d_\lambda(n) = (-1)^m d_{{\lambda^T}}(-n)\,,
\end{equation}
for a Young diagram with $m$ boxes.

As a first example, we consider the bilinear combination $M_{ab} \phi^a \phi^b$. In an $O(2N)$-symmetric theory, the physical degrees of freedom of $M_{ab}$ are $2N(2N+1)/2$, as only the symmetric part of is independent when commuting the fields.
On the other hand, whenever we consider an $Sp(2N)$-invariant theory built from anticommuting fields, the physical degrees of freedom of the matrix $M_{ab}$ are $2N(2N-1)/2$, since only the antisymmetric part survives. 
When we decompose this operator into irreducible representations, we 
subtract the trace of $M_{ab}$.
The degrees of freedom of the traceless symmetric irrep are therefore $2N(2N+1)/2 -1$. 
To keep the correspondence between the theories valid, we have to define a traceless condition for operators in $Sp(2N)$. 
This condition is realized by contracting the indices with the symplectic matrix \eqref{eq:symp_J},
\begin{equation}
M_{ab}J^{ab} = 0\,.
\end{equation}
This provides the matching condition to subtract the trace degrees of freedom.

As a second example, we note that the symmetric part of $N_{ab}$ leads to a total derivative in 
the bilinear combination of commuting scalars $N_{ab} \partial_\mu \phi^a \phi^b$. 
The physical degrees of freedom of $N_{ab}$ are therefore $2N(2N-1)/2$ since only the antisymmetric part survives. The same is true in an anticommuting theory, but again exchanging the role of symmetrization and antisymmetrization of indices, as the map between the theories prescribes. In this case, there is no trace part to subtract.

The trace subtraction is important when considering the correct map among larger representations. 
Tab.~\ref{Tab:multiplicities} lists some of the dimensions of the irreducible representations of $O(n)$.\footnote{In this paragraph, for simplicity, we take $n$ even and avoid writing $2N$.}
If we try to map the dimensions of the representations by changing the sign of $n$, we immediately see the map does not work: the reason is precisely the subtraction of traces. 
Consider, for example, the $T_4$ representation. 
The dimension of a rank-$4$ symmetric tensor is $\binom{n+3}{4}$. 
The contraction with $\delta_{ab}$ gives a rank-$2$ symmetric tensor, whose dimension is $\binom{n+1}{2}$. 
Overall, we find
\begin{equation}
d_{T_4,O(n)}=\binom{n+3}{4} - \binom{n+1}{2} = \frac{n(n-1)(n+1)(n+6)}{24}\,.
\end{equation}
To find the dimension of the corresponding irrep in $Sp(n)$, we consider the transpose Young tableau, which gives the representation $A_4$, and we subtract the dimension of the irrep $A$ corresponding to the tracelessness condition giving an antisymmetric rank-$2$ tensor. 
The result is
\begin{equation}
d_{A_4,Sp(n)}= \binom{n}{4} - \binom{n}{2} = \frac{n(n+1)(n-1)(n-6)}{24}\,,
\end{equation}
matching the result above under the map $n\rightarrow -n$. 
The same can procedure be performed for other Young diagrams, for example for the $B_4$ representation. 
In $O(n)$, the trace subtraction leads to
\begin{equation}
d_{B_4,O(n)}=\frac{n^2(n^2-1)}{12} - \binom{n+1}{2} = \frac{n(n+1)(n+2)(n-3)}{12}\,.
\end{equation}
The same calculation for $Sp(n)$,
where the corresponding representation has the same Young tableau but a different trace subtraction, results in
\begin{equation}
d_{B_4,Sp(n)}=\frac{n^2(n^2-1)}{12} - \binom{n}{2} = \frac{n(n-1)(n-2)(n+3)}{12}\,,
\end{equation}
as expected.

\subsection[Spectrum constraints in the \texorpdfstring{$Sp(n)$}{Sp(n)} model]
{\boldmath Spectrum constraints in the $Sp(n)$ model}
\label{sec:constraintsSp}

One of the consequences we draw from this argument is that, when looking at Table~\ref{Tab:multiplicities}, we can make sense of the representations at negative even values of $n$ by considering operators in the anticommuting $Sp(n)$ theory. When substituting actual values for $n$, we can observe that the dimensions of some of the representations becomes negative, which is a useful diagnostic for which operators are subject to spectrum constraints. However, we note that matching dimensions alone does generally not completely identify annihilation partners; the character-specialization criterion of Ref.~\cite{Cao:2023psi} supplies a stronger condition. 
Characters of irreducible $Sp(n)$ representations can likewise be obtained in terms of symplectic Schur functions \cite{Littlewood1955,KoikeTerada1987,ALBION2024104000}.

In the examples below, the resulting constraints on anomalous dimensions are illustrated by the intersections in Fig.~\ref{fig:examplesDim6} at negative $n$.
The fact that the correspondence between the two theories is only possible for an even number $n$ suggests that these intersections occur only for even values; all our plots and explicit results corroborate this statement. 
Negative odd values are possible in the $O(n-m)$ framework proposed in \cite{McKane:1979rm}, which includes $n$ commuting fields and $m$ anticommuting fields. We leave the analysis of possible constraints arising in such theories to further work.

\paragraph{\boldmath An explicit example: the operator $\phi^6$.}
Let us further illustrate the map between operators in the $O(n)$ and $Sp(n)$ theories and the spectrum constraints at negative $n$ with a simple example. We consider an operator made of six (commuting) scalars,
\begin{equation}\label{opeq6}
\mathcal{O}_{\phi^6} =  \frac{c_{\phi^6}^{abcdef}}{6!} \phi^a 
    \phi^b\phi^c\phi^d\phi^e\phi^f\,,
\end{equation}
where $c_{\phi^6}^{abcdef}$ is fully symmetric. This tensor can be decomposed into four irreducible representations of $O(n)$: the totally traceless symmetric representation $T_6$, the four-index traceless symmetric $T_4$, the traceless symmetric $T$ and the singlet $S$,
\begin{equation}
 \raisebox{0.5mm}{\scalebox{0.35}{$\ydiagram{6}$}}_{\,S_6} \simeq \   
 \raisebox{0.5mm}{\scalebox{0.35}{$\ydiagram{6}$}}_{\,O(2N)} +
 \raisebox{0.5mm}{\scalebox{0.35}{$\ydiagram{4}$}}_{\,O(2N)} +
 \raisebox{0.5mm}{\scalebox{0.35}{$\ydiagram{2}$}}_{\,O(2N)} +
 S_{\,O(2N)}
\end{equation}
(using the notation introduced in Sec.~\ref{sec:2.1RepTheory}).
We have previously established that for (integer) negative values of $N$, the operators can be defined in an $Sp(2N)$-invariant theory with anticommuting scalars and positive $N$.  The six-index symmetric traceless operator labeled by the representation $T_6$ of the $O(2N)$ group corresponds, for negative $N$, to the antisymmetric representation to which the $Sp(2N)$ traces have been removed. 
The same happens for the $T_4$ representation, from which we subtract the $n(n-1)/2$ contractions after transposing the Young tableau; and to the $T_2$ representation from which we remove one antisymmetric trace.

We can count the degrees of freedom of the representations to predict which $\phi^6$ operators are going to annihilate for different values of $N$, in a similar fashion to the analysis in Sec.~\ref{sec:inv}. The results are the intersections shown in Fig.~\ref{fig:examplesDim6} (using shorthand $\dot c = dc/d\log\mu$),
{\allowdisplaybreaks
\begin{align}
    &n=0:& \dot c^{(S)}_{\phi^6} = \dot c^{(T_2)}_{\phi^6}\,, \nn\\
    &n=-2:& \dot c^{(S)}_{\phi^6} = \dot c^{(T_4)}_{\phi^6}\,, \nn\\
    &n=-4:& \dot c^{(S)}_{\phi^6} = \dot c^{(T_6)}_{\phi^6}\,,
    && \dot c^{(T)}_{\phi^6} = \dot c^{(T_4)}_{\phi^6}
    \,,\nn\\
    &n=-6:& \dot c^{(T)}_{\phi^6} = \dot c^{(T_6)}_{\phi^6}\,,\nn\\
    &n=-8:& \dot c^{(T_4)}_{\phi^6} = \dot c^{(T_6)}_{\phi^6}\,.\label{eq3.18}
\end{align}
Consider} for example $n=-4$, where two crossings occur: the negative-dimensional representations are $T_6$ and $T_4$, which annihilate with, respectively the representations $S$ and $T_2$. 
This pairing simply follows from matching the dimensions (or more generally the characters): for $n=-4$, the dimensions of 
$T_6$ and S are $-1$ and $1$, respectively, while  those of $T_4$ and $T$ are $-5$ and $5$, respectively.
Note that this argument can straightforwardly be generalized to other operators;
we derive the spectrum constraints between all traceless symmetric operators in Sec.~\ref{Sec:family}.

\subsection
  [Perturbative derivation of the spectrum constraints for any \texorpdfstring{$n$}{n}]
  {\boldmath Perturbative derivation of the spectrum constraints for any $n$}
\label{sec:pertProof}

To complement the previous discussion on the spectrum constraints for negative $n$, we will present an alternative derivation in this subsection, which does not rely on the character continuation of Ref.~\cite{Cao:2023psi} nor on the duality between $O(-2N)$ and $Sp(2N)$. That is, we will assume $n$ to be an integer that is large enough to avoid low-rank degeneracies, and we will work with explicit tensor structures as in Sec.~\ref{sec:inv}, without doing any loop calculations.

Let us start by rederiving the constraints on the anomalous dimension of Lorentz-singlet dimension-six operators. The two operators with general flavor structure were given in Eqs.~\eqref{opeq2.1} and \eqref{opeq6}, which have permutation structure given by \,$\raisebox{1.2mm}{\scalebox{0.35}{$\ydiagram{2,2}$}}_{\,S_4}$
and $\raisebox{0.5mm}{\scalebox{0.35}{$\ydiagram{6}$}}_{S_6}$, respectively.
The anomalous dimensions of the coupling constant tensors of these operators in the $O(n)$ model are given in full generality by
{
\begin{align}
    \frac{d\,c_{\phi^6}^{abcdef}}{d\log\mu} &=
        f_{6,1}\,c_{\phi^6}^{abcdef} + f_{6,2}\left(
        c_{\phi^6}^{abcdgg}\delta^{ef} + 14 \text{ permutations}
        \right)\nn\\&\quad 
        + f_{6,3}\left(
        c_{\phi^6}^{abgghh}(\delta^{cd}\delta^{ef}+\delta^{ce}\delta^{df}+\delta^{cf}\delta^{de})
        + 14 \text{ permutations}
        \right)\nn\\&\quad 
        + f_{6,4}\left(c_{\phi^6}^{gghhii}\delta^{ab}\delta^{cd}\delta^{ef}+
        14 \text{ permutations}\right)\nn\\&\quad
        + f_{6,5}\left(c_{\phi^4\square}^{abgg}(\delta^{cd}\delta^{ef}+\delta^{ce}\delta^{df}+\delta^{cf}\delta^{de}) + 14 \text{ permutations} \right)\nn\\&\quad
        +f_{6,6} \left( c_{\phi^4\square}^{gghh}\delta^{ab}\delta^{cd}\delta^{ef}+
        14 \text{ permutations}\right),
        \label{3.3}
    \displaybreak[3]\\
    \frac{d\,c_{\phi^4\square}^{abcd}}{d\log\mu} &=
    f_{4,1} \, c^{abcd}_{\phi^4\square}
    \nn\\&\quad+f_{4,2} \left(c^{abee}_{\phi^4\square}\delta^{cd}
+ c^{cdee}\delta^{ab}
-\tfrac12 c^{acee}\delta^{bd}
-\tfrac12 c^{bdee}\delta^{ac}
-\tfrac12 c^{adee}\delta^{bc}
-\tfrac12 c^{bcee}\delta^{ad}
    \right)\nn\\&\quad
+f_{4,3} \, c_{\phi^4\square}^{eeff}\left(\delta^{ab}\delta^{cd} - \tfrac{1}{2}\delta^{ac}\delta^{bd}-\tfrac12\delta^{ad}\delta^{bc}\right)\nn\\&\hspace{-8mm}
+f_{4,4}\left(c_{\phi^6}^{abeeff}\delta^{cd}+c_{\phi^6}^{cdeeff}\delta^{ab}
-\tfrac12 c_{\phi^6}^{aceeff}\delta^{bd}
-\tfrac12 c_{\phi^6}^{bdeeff}\delta^{ac}
-\tfrac12 c_{\phi^6}^{adeeff}\delta^{bc}
-\tfrac12 c_{\phi^6}^{bceeff}\delta^{ad}\right)\nn\\&\quad
+f_{4,5} \, c^{eeffgg}_{\phi^6}\left(\delta^{ab}\delta^{cd} - \tfrac{1}{2}\delta^{ac}\delta^{bd}-\tfrac12\delta^{ad}\delta^{bc}\right),
\end{align}
where} the renormalization group equations inherit the index-permutation symmetries of the corresponding coupling on the left-hand side
In perturbation theory, the $f_{6,i}=f_{6,i}(\lambda,n)$ and $f_{4,i}=f_{4,i}(\lambda,n)$ are polynomials in the coupling constant and the number of flavors. 

Decomposing the anomalous dimensions of the reducible representations into the corresponding irreps, we find {\allowdisplaybreaks
\begin{align}
    \frac{d\,c^{(S)}_{\phi^6}}{d\log\mu}  & = 
    c^{(S)}_{\phi^6} \bigg[ f_{6,1}
    +3\,(n+4)f_{6,2}
    +3\,(n+2)(n+4)f_{6,3}
    +n(n+2)(n+4)f_{6,4}
    \bigg]\nn\\&\quad
    +c^{(S)}_{\phi^4\square}\,(n-1)\,\bigg[
    6\,f_{6,5}+2n\,f_{6,6}\bigg]\nn\\
    \frac{d\,c^{(T)}_{\phi^6}}{d\log\mu}  & = 
    c^{(T)}_{\phi^6} \bigg[ f_{6,1}
    +2\,(n+6)f_{6,2}
    +(n+4)(n+6)f_{6,3}
    \bigg]
    +\tfrac12 c^{(T)}_{\phi^4\square}\,(n-2)\,f_{6,5}
    \nn\\
    \frac{d\,c^{(T_4)}_{\phi^6}}{d\log\mu}  & = 
    c^{(T_4)}_{\phi^6} \bigg[ f_{6,1}
    +(n+8)f_{6,2}
    \bigg]
    \nn\\
    \frac{d\,c^{(T_6)}_{\phi^6}}{d\log\mu}  & = 
    c^{(T_6)}_{\phi^6} \, f_{6,1}
    \nn\\[2mm]
    \frac{d\,c^{(S)}_{\phi^4\square}}{d\log\mu}  & = 
    c^{(S)}_{\phi^4\square} \bigg[ f_{4,1}
    +2\,(n-1)f_{4,2}
    +n(n-1)f_{4,3}
    \bigg]
    +c^{(S)}_{\phi^6}\,(n+2)(n+4)\,\bigg[
    f_{4,4}+\tfrac12 n\,f_{4,5}\bigg]\nn\\
    \frac{d\,c^{(T)}_{\phi^4\square}}{d\log\mu}  & = 
    c^{(T)}_{\phi^4\square} \bigg[ f_{4,1}
    +(n-2)f_{4,2}
    \bigg]
    +c^{(T)}_{\phi^6}\,2(n+4)(n+6)\,
    f_{4,4}\nn\\
    \frac{d\,c^{(B_4)}_{\phi^4\square}}{d\log\mu}  & = 
    c^{(B_4)}_{\phi^4\square} \, f_{4,1}
    \label{eq:fdim6}
\end{align}
from} which we can directly read off the spectrum constraints that were previously derived in Sections~\ref{sec:review} and~\ref{sec:constraintsSp},
without any knowledge on the specific forms of $f_{6,i}$ and $f_{4,i}$ (apart from them being regular at the integer $n$ of interest), which would need to be determined by an explicit computation.
These reproduce all spectrum constraints that are visible as intersections in Fig.~\ref{fig:examplesDim6}. 

As in the explicit results in Eq.~\eqref{2.17}, it is relevant to note that the $S$ and $T$ representations have two operators at this order, so in general one needs to diagonalize the renormalization group mixing matrix to extract the scaling dimensions. For most of the relations involving $S$ or $T$ above, however, one of the off-diagonal entries is zero, so the eigenvalues correspond to the diagonal entries directly---this is not a coincidence, as we explain in Sec.~\ref{sec:3}. The only exception is for constraints that involve two sets of operators that mix:
\begin{align}
\frac{d}{d\log\mu}
    \begin{pmatrix}
     c^{(S)}_{\phi^6}\\[2mm]  c^{(S)}_{\phi^4\square}
    \end{pmatrix}\bigg|_{n=0}&=
    \begin{pmatrix}
        f_{6,1}
    +12\,f_{6,2}
    +24\,f_{6,3}
    &
    -6\,f_{6,5}\\[2mm]
    8\,f_{4,4}
    &
    f_{4,1}
    -2f_{4,2}
    \end{pmatrix},\nn\\[2mm]
   \frac{d}{d\log\mu}
    \begin{pmatrix}
     c^{(T)}_{\phi^6}\\[2mm]  c^{(T)}_{\phi^4\square}
    \end{pmatrix}\bigg|_{n=0}&=
    \begin{pmatrix}
        f_{6,1}
    +12\,f_{6,2}
    +24\,f_{6,3}
    &
    -f_{6,5}\\[2mm]
    48\,f_{4,4}
    &
    f_{4,1}
    -2f_{4,2}
    \end{pmatrix},\label{eq:2.7}
\end{align}
which have the same eigenvalues. These matrices could be made to agree exactly by changing the overall normalizations of the operators or tensor structures.

In this derivation, we have parametrized the anomalous dimensions of seven operators in terms of seven unknown functions $f_{\{4,6\},i}$, in such a way that the spectrum constraints become manifest. In general, the number of unknowns will be the same as the number of independent anomalous dimensions. However, in Sec.~\ref{Sec:family} we will argue that some of the $f_{\{4,6\},i}$ vanish at low orders in perturbation theory. This results in an overconstrained system, which makes it possible to `bootstrap' the scaling dimensions of some operators in terms of others. We use this to derive new anomalous dimensions from existing results without additional loop calculations.

\begin{figure}[t]
\includegraphics[width=\textwidth, trim=0 0.4cm 0 0, clip]{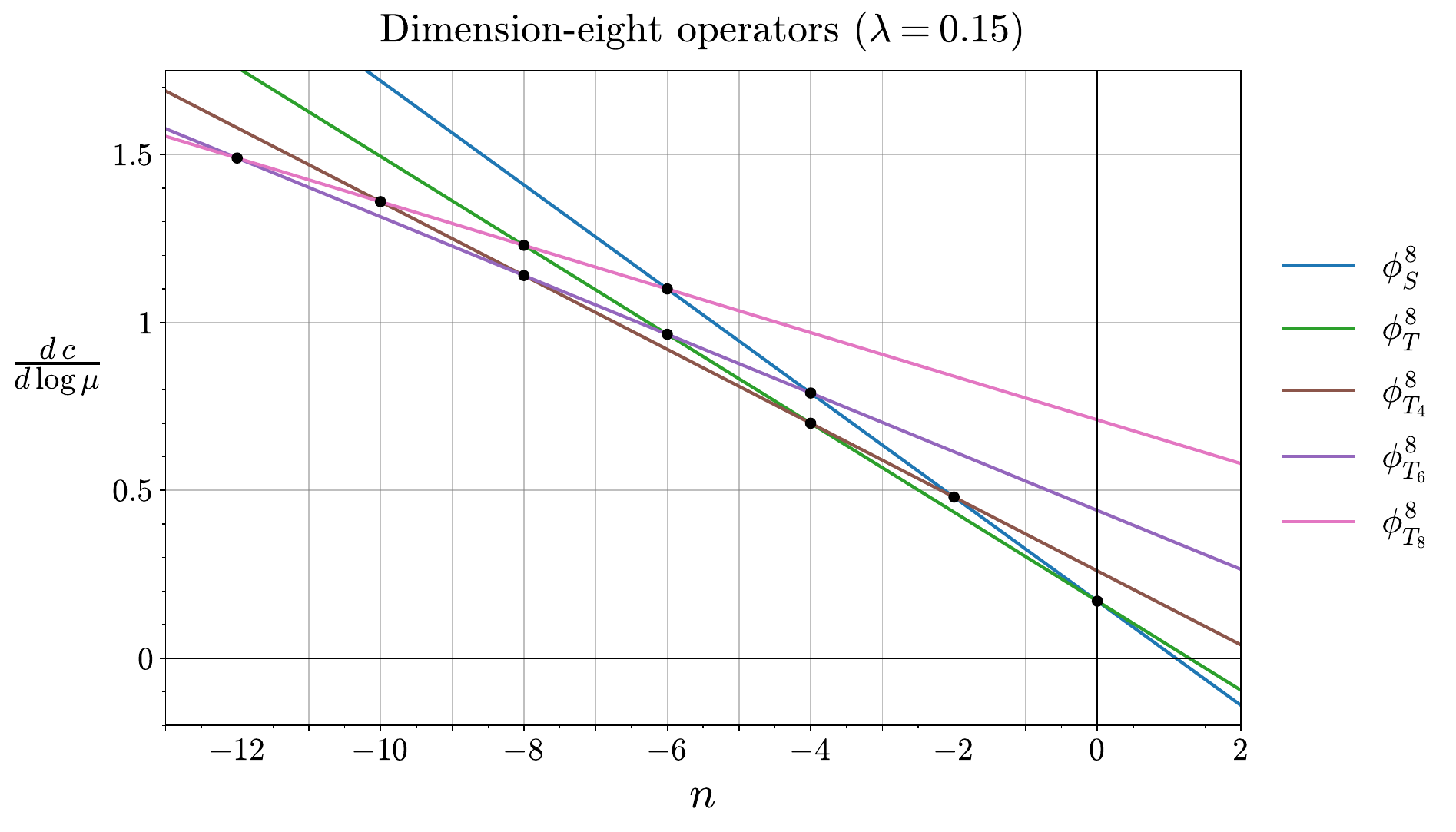}
\hspace{-5mm}
\caption{
Anomalous dimensions of the Wilson coefficients of dimension-eight operators in the $O(n)$ model at two loops as a function of $n$~\cite{Henriksson:2025hwi,Henriksson:2025vyi}.
This plot includes only the operators with field content $\phi^8$ in the free theory limit. 
The marginal $\phi^4$ coupling is chosen to be $\lambda=0.15$ for optimal illustration of the spectrum constraints (which hold for any $\lambda$). }
\label{fig:examplesDim8}
\end{figure}

\paragraph{Dimension eight.} The same approach can be applied at dimension eight, for example to the operators of the form 
\begin{equation}
    \phi^a\phi^b\phi^c\phi^d\phi^e\phi^f\phi^g\phi^h
    \quad \leftrightarrow \quad 
    \raisebox{0.5mm}{\scalebox{0.35}{$\ydiagram{8}$}}_{S_8}\,.
\end{equation}
Up to two loops, these operators do not mix with operators with derivatives. The anomalous dimension tensor then takes the form
\begin{align}
    \frac{d\,c_{\phi^8}^{abcdefgh}}{d\log\mu} &=
    f_{8,1}\,c_{\phi^8}^{abcdefgh} + 
    f_{8,2}\left(c_{\phi^8}^{abcdefii}\delta^{gh} + 
    27 \text{ permutations} \right)\nn\\&\quad
    + 
    f_{8,3}\left(c_{\phi^8}^{abcdiijj}\delta^{ef}\delta^{gh} + 
    209 \text{ permutations} \right)\nn\\&\quad
    + 
    f_{8,4}\left(c_{\phi^8}^{abiijjkk}\delta^{cd}\delta^{ef}\delta^{gh} + 
    419 \text{ permutations} \right)\nn\\&\quad
    + 
    f_{8,5}\left(c_{\phi^8}^{iijjkkll}\delta^{ab}\delta^{cd}\delta^{ef}\delta^{gh} + 
    104 \text{ permutations} \right),
    \label{3.8}
\end{align}
leading to {
\begin{align}
    \frac{d\,c_{\phi^8}^{(S)}}{d\log\mu} &=
    c_{\phi^8}^{(S)}\left[f_{8,1}+(n+6)
    \bigg(
    4\,f_{8,2}
    +(n+4)\Big(6\,f_{8,3}+(n+2)\big(4\,
    f_{8,4}+n\,f_{8,5}\big)\Big)\bigg)\right],\nn\\
    \frac{d\,c_{\phi^8}^{(T)}}{d\log\mu} &=
    c_{\phi^8}^{(T)}\left[f_{8,1}+(n+8)\Big(3\,f_{8,2}+(n+6)\big(3\,f_{8,3}+(n+4)f_{8,4}\big)\Big)\right],\nn\\
    \frac{d\,c_{\phi^8}^{(T_4)}}{d\log\mu} &=
    c_{\phi^8}^{(T_4)}\left[f_{8,1}+(n+10)\big(2\,f_{8,2}+(n+8)f_{8,3}\big)
    \right],\nn\\
    \frac{d\,c_{\phi^8}^{(T_6)}}{d\log\mu} &=
    c_{\phi^8}^{(T_6)}\left[ f_{8,1}+(n+12)f_{8,2}\right],
    \nn\\
    \frac{d\,c_{\phi^8}^{(T_8)}}{d\log\mu} &=
    c_{\phi^8}^{(T_8)}\,f_{8,1}\,,
\end{align}
from} which the spectrum constraints can directly be read off. The two-loop results computed in~\cite{Henriksson:2025hwi,Henriksson:2025vyi} for all representations have been plotted in Fig.~\ref{fig:examplesDim8} to visualize all spectrum constraints, where we highlight that all of them occur for $n\leq0$. In the next section, we will extend this result to arbitrary $\phi^k$-type operators, deriving a closed form of the spectrum constraints for any representation.

\section{New perturbative results for a family of operators}\label{Sec:family}

In the previous section, we discussed the spectrum constraints in the $O(n)$ model for negative $n$. The plots of the anomalous dimensions as a function of $n$, Figs.~\ref{fig:examplesDim6} and \ref{fig:examplesDim8}, suggest rigid structures in which the dimension of one operator may be fixed in terms of others by imposing consistency with the large number of intersection points.
To illustrate this possibility, we will exploit the spectrum constraints in this section to derive new perturbative results for the scaling dimensions of a particular class of operators without calculating additional diagrams. 
We will focus on the renormalization group equations of $\phi^k$-type operators, which include $\lfloor\tfrac{k}{2}\rfloor+1$ irreps $T_k$, $T_{k-2}$, etc.\ (where $T_1\equiv V$ and $T_0\equiv S$), 
and the spectrum constraints require every pair of scaling dimensions to coincide at some $n\leq0$; see Fig.~\ref{fig:examplesDim8} for the $\phi^8$-type operators. Moreover, we will work up to two loops, such that there is no mixing with operators with derivatives~\cite{Bern:2019wie}. 

The renormalization group equations for $c_{\phi^6}^{abcdef}$ in Eq.~\eqref{3.3} and for $c_{\phi^8}^{abcdefgh}$ in Eq.~\eqref{3.8} can be generalized to operators with an arbitrary number of fields,
\begin{align}
\mathcal{O}_{\phi^k} &= 
c_{\phi^k}^{a_1\dots a_k}\,\phi^{a_1}\cdots \phi^{a_k}\,,\\
\frac{d\,c_{\phi^k}^{a_1 \dots a_k}}{d\log\mu} &= f_1(\lambda,k,n)\,c_{\phi^k}^{a_1\dots a_k} 
+ f_2(\lambda,k,n)
\Big(\delta^{a_1a_2}c_{\phi^k}^{bba_3\dots a_k} + \text{perms}\Big)
+ \dots\,, \label{phikGen}
\end{align}
where the total number of permutations in the second term is $\tfrac12 k(k-1)$ and we have suppressed terms with at least two Kronecker delta functions in which the tensor $c_{\phi^k}$ has at least four contracted indices. 
In the following, we will argue that the suppressed terms in Eq.~\eqref{phikGen} 
are zero at one and two loops. To see this, 
first note that the only one-loop topology is
\begin{align}
\begin{tikzpicture}[scale=0.7,baseline={(L.center)}]
\begin{feynman}
\vertex (L) at (0,0);
\vertex (R) at (1.1,0);
\vertex (a1) at (-0.7,0.45);
\vertex (a4) at (-0.7,-0.45);
\vertex (b1) at (1.8,0.25);
\vertex (b2) at (1.8,-0.25);
\diagram*{
(a1) -- (L),
(a4) -- (L),
(L) -- [half left,looseness=1.2] (R),
(L) -- [half right,looseness=1.2] (R),
(R) -- (b1),
(R) -- (b2),
};
\node at (-0.6,0.15) {\scriptsize$\vdots$};
\draw[decorate,decoration={brace,mirror,amplitude=3pt}] (-0.95,0.52) -- (-0.95,-0.52) node[midway,left=4pt] {$k-2$};
\end{feynman}
\end{tikzpicture} \ ,
\end{align}
because we allow for a single insertion of the $\phi^k$ operator and restrict the remaining vertices to $\phi^4$. This means that at one loop we may only contract up to two indices of $c_{\phi^k}^{a_1\dots a_k}$. 
Similarly, the following two-loop topologies involve at most two contracted indices of $c_{\phi^k}^{a_1\dots a_k}$,
\begin{align}
\begin{tikzpicture}[scale=0.7,baseline={(L.center)}]
\begin{feynman}
\vertex (L) at (0,0);
\vertex (U) at (1.0,0.4);
\vertex (D) at (1.0,-0.4);
\vertex (a1) at (-0.7,0.45);
\vertex (a2) at (-0.7,-0.45);
\vertex (b1) at (1.7,0.6);
\vertex (b2) at (1.7,-0.6);
\diagram*{
(a1) -- (L),
(a2) -- (L),
(L) -- (U),
(L) -- (D),
(U) -- [bend right=45] (D),
(U) -- [bend left=45] (D),
(U) -- (b1),
(D) -- (b2),
};
\node at (-0.6,0.15) {\scriptsize$\vdots$};
\draw[decorate,decoration={brace,mirror,amplitude=3pt}] (-0.95,0.52) -- (-0.95,-0.52) node[midway,left=4pt] {$k-2$};
\end{feynman}
\end{tikzpicture}\ , \quad 
\begin{tikzpicture}[scale=0.7,baseline={(L.center)}]
\begin{feynman}
\vertex (L) at (0,0);
\vertex (R) at (1.1,0);
\vertex (V) at (1.55,0);
\vertex (a1) at (-0.7,0.45);
\vertex (a2) at (-0.7,-0.45);
\vertex (b1) at (2.15,0);
\vertex (c1) at (1.75,0.35);
\vertex (c2) at (1.75,-0.35);
\diagram*{
(a1) -- (L),
(a2) -- (L),
(L) -- (R),
(L) -- [half left,looseness=1.35] (R),
(L) -- [half right,looseness=1.35] (R),
(R) -- (V),
(V) -- (b1),
(V) -- (c1),
(V) -- (c2),
};
\node at (-0.6,0.15) {\scriptsize$\vdots$};
\draw[decorate,decoration={brace,mirror,amplitude=3pt}] (-0.95,0.52) -- (-0.95,-0.52) node[midway,left=4pt] {$k-3$};
\end{feynman}
\end{tikzpicture}\ , \quad
\begin{tikzpicture}[scale=0.7,baseline={(L.center)}]
\begin{feynman}
\vertex (L) at (0,0);
\vertex (M) at (1.0,0);
\vertex (R) at (2.0,0);
\vertex (a1) at (-0.7,0.45);
\vertex (a2) at (-0.7,-0.45);
\vertex (b1) at (2.7,0.25);
\vertex (b2) at (2.7,-0.25);
\diagram*{
(a1) -- (L),
(a2) -- (L),
(L) -- [half left,looseness=1.2] (M),
(L) -- [half right,looseness=1.2] (M),
(M) -- [half left,looseness=1.2] (R),
(M) -- [half right,looseness=1.2] (R),
(R) -- (b1),
(R) -- (b2),
};
\node at (-0.6,0.15) {\scriptsize$\vdots$};
\draw[decorate,decoration={brace,mirror,amplitude=3pt}] (-0.95,0.52) -- (-0.95,-0.52) node[midway,left=4pt] {$k-2$};
\end{feynman}
\end{tikzpicture}\ .
\quad
\end{align}
(For the middle diagram, we note that it is impossible to contract three indices with Kronecker delta functions.)
Finally, the only remaining two-loop topology is
\begin{align}
\begin{tikzpicture}[scale=0.7,baseline={(L.center)}]
\begin{feynman}
\vertex (L) at (0,0);
\vertex (M) at (1.0,0);
\vertex (R) at (2.0,0);
\vertex (a1) at (-0.7,0.25);
\vertex (a2) at (-0.7,-0.25);
\vertex (b1) at (2.7,0.25);
\vertex (b2) at (2.7,-0.25);
\vertex (c1) at (0.7,0.8);
\vertex (c2) at (1.3,0.8);
\diagram*{
(a1) -- (L),
(a2) -- (L),
(L) -- [half left,looseness=1.2] (M),
(L) -- [half right,looseness=1.2] (M),
(M) -- [half left,looseness=1.2] (R),
(M) -- [half right,looseness=1.2] (R),
(R) -- (b1),
(R) -- (b2),
(c1) -- (M),
(c2) -- (M),
};
\node at (1.0,0.7) {\scriptsize$\cdot \, \cdot$};
\draw[decorate,decoration={brace,amplitude=3pt}] (0.62,0.95) -- (1.38,0.95) node[midway,above=4pt] {$k-4$};
\end{feynman}
\end{tikzpicture} \ ,\label{diag4.5}
\end{align}
which does involve the contraction of four indices of $c_{\phi^k}^{a_1\dots a_k}$. 
However, importantly, the associated diagram factorizes into two one-loop subdiagrams on the middle vertex. 
As was shown in Ref.~\cite[Sec.\,5]{Jenkins:2023rtg}, 
the overall UV divergence of such a diagram may contain an $1/\varepsilon^2$ pole, but no $1/\varepsilon$ pole in the 
minimal subtraction scheme.%
    \footnote{As explained in Ref.~\cite[Sec.\,5]{Jenkins:2023rtg}, this conclusion may be different if the vertices carry momenta. This is relevant in the next section where we consider $\phi^6\square^2$ operators with four derivatives.} 
Hence, the diagram \eqref{diag4.5} does not contribute to the anomalous dimensions in this scheme.
This concludes the argument that only $f_1$ and $f_2$ in \eqref{phikGen} are nonzero up to two loops, such that the suppressed terms are necessarily of order $\lambda^3$.

While generally the scaling dimensions of the 
$\lfloor\tfrac{k}{2}\rfloor+1$ $\phi^k$-type operators are determined by the same number of unknowns as operators, we thus find that up to two loops these are fully determined by only two polynomials in $\lambda$ and $n$: $f_1$ and $f_2$. 
The general operator $\mathcal{O}_{\phi^k}$ can be mapped to the $m$-index traceless symmetric representations of the $O(n)$ global symmetry (for $k-m$ an even integer) by replacing $c^{a_1\dots a_k}\to (t^{a_1}\cdots t^{a_m} \delta^{a_{m+1}a_{m+2}}\cdots \delta^{a_{k-1}a_k} + \text{perms}$, where the product of $t^a$ vectors with $t^2=0$ projects to the traceless symmetric representation, 
\begin{align}
    \mathcal{O}_{\phi^k,T_m} = 
    c_{\phi^k,T_m}\,
    (t\cdot \phi)^m\,
    \phi^{a_1}\cdots \phi^{a_m} \, 
    (\phi\cdot\phi)^{(k-m)/2}\,,
\end{align}
with scaling dimension (using $\Delta_{\phi^k,T_m} = k(1-\varepsilon/2) + 1/c_{\phi^k,T_m} \cdot d c_{\phi^k,T_m}/d\log\mu$)
\begin{align}
\Delta_{\phi^k,T_m} &= 
k\,(1-\tfrac{\varepsilon}{2})+
f_1(\lambda,k,n)
+ \tfrac12(k-m)(n+k+m-2)\,
f_2(\lambda,k,n)
+ O(\lambda^3)\,.\label{scalingDimGenk}
\end{align}
When treating this scaling dimension as a function of $n$, it follows that
\begin{align}
\Delta_{\phi^k,T_{m'}}(n) = \Delta_{\phi^k,T_m}(n)\,,
\quad \text{when } n = 2-m-m' \text{ (or $m=m'$)}\,,
\end{align}
which captures the spectrum constraints between all $m$-index traceless symmetric representations.
Although we suppressed $O(\lambda^3)$ terms in the derivation, we expect this constraint to hold at all orders in $\lambda$.

We will now bootstrap the scaling dimension \eqref{scalingDimGenk} for general $k$ and $m$ up to two loops.
The scaling dimension $\Delta_{\phi^k,T_m}$ is known at one loop for any value of $k$ and 
$m$~\cite{Wegner:1972zz,Henriksson:2022rnm},%
\footnote{Ref.~\cite{Henriksson:2022rnm} provides the scaling dimensions at the conformal fixed point in $d=4-\varepsilon$ dimensions. Here we write the expression in terms of $\lambda = 3\varepsilon/(n+8)+9(3n+14)\varepsilon^2/(n+8)^3$, which can then be continued away from the fixed point.}
\begin{align}
    \Delta_{\phi^k,T_{m}} =
    k\,(1-\tfrac{\varepsilon}{2}) + 
    \tfrac16 \left(
    3k^2+k(n-4)-m(n+m-2)
    \right)\lambda + O(\lambda^2)\,.
\end{align}
from which we could extract $f_1$ and $f_2$ directly at order $\lambda$.
In fact, 
$\Delta_{\phi^k,T_m}$ is known at six loops for the completely traceless symmetric representation ($m=k$)~\cite{Bednyakov:2022guj}, which determines $f_1$ up to order $\lambda^6$. 
Since $\Delta_{\phi^k,T_m}$ is additionally known 
at two loops for the singlet representation ($m=0$)~\cite{Derkachov:1997gc,Henriksson:2022rnm}, we can exploit the spectrum constraints to obtain the scaling dimensions for all $\phi^k$ operators at two loops, without additional loop calculations.%
    \footnote{Results are also available for the $\phi^k$ operators in the singlet~\cite{Bednyakov:2025usv} and the maximally traceless-symmetric representation~\cite{Jack:2020wvs} in the three-dimensional $O(n)$ model, which allows to repeat our analysis in $d=3-\varepsilon$ dimensions.}
This results in
\begin{align}
    f_1(\lambda,k,n) &= \frac{k(k-1)\lambda}{3}
    - \frac{k (2 + 8 k^2 + 2 k (n-6) - 3 n)\lambda^2}{36}+O(\lambda^3)\,,\nn\\
    f_2(\lambda,k,n) &= \frac{\lambda}{3} 
    - \left(\frac{k}{2}-\frac{7}{9}\right)\lambda^2 + O(\lambda^3)\,,
\end{align}
which to the best of our knowledge was not known in the literature.
It is useful to compare our expression with the expected large-$k$ behavior from a semi-classical calculation~\cite{Antipin:2025ilv},
which for instance implies that $f_2$ 
cannot depend on $k$ at one loop because its coefficient in~\eqref{scalingDimGenk}
already goes like $\sim k^2$ for large $k$. Similarly, $f_2$ does not depend on $n$ at one and two loops because 
an $n$-dependent contribution to $f_2$ would generate terms proportional to $n^2$, and
three loops are required to obtain such terms. (The only topology that would result in factors of $n^2$ would be \eqref{diag4.5}, but this diagram does not contribute to the anomalous dimensions.)

\paragraph{Three-loop extension.}
It is not straightforward to extend the above analysis to obtain a three-loop expression for $\Delta_{\phi^k,T_{m}}$ without additional computations. First, an additional data point would be required to determine $f_3$ which multiplies $(\delta^{a_1 a_2}\delta^{a_3 a_4} c^{b_1b_1b_2b_2a_5\dots a_k} + \text{perms})$. Second, while $\phi^k$ does not mix with operators with derivatives up to two loops~\cite{Bern:2019wie}, there may be such mixing at three loops. Diagonalizing this mixing problem may result in non-polynomial dependence on $k$, $m$ and $n$, which requires a modification of the above argument.

\section{Operator mixing and non-renormalization}\label{sec:3}

Up to this point, we have focused on the operator spectrum, given by the eigenvalues of the renormalization-group mixing matrices. We now turn to the constraints on operator mixing itself that arise when states leave the spectrum at specific values of $n$.
While individual entries in mixing matrices are necessarily basis dependent, the organization of the theory in terms of $S_m$ representations in Eq.~\eqref{eq:ONLagrGen} provides a natural (and preferred) basis in which to expose the structure due to evanescence.
We will focus on singlet operators. Importantly, the annihilation of singlet operators by operators in other irreps is reflected in the mixing between singlet operators only. This is relevant in EFTs with an exact symmetry, in which one only includes higher-dimensional operators in the singlet representation of a global symmetry. In Sec.~\ref{sec:n=-2}, we discuss additional constraints on the mixing matrix and its eigenvalues in the $O(n)$ model that occur for $n=-2$.

\subsection{Mixing and non-renormalization of singlet operators}\label{sec:5.1}

\paragraph{Dimension six.}
Focusing on the operator mixing between Lorentz-scalar $O(n)$-singlet operators at dimension six, Eq.~\eqref{eq:fdim6} implies the renormalization group equation%
    \footnote{The overall factor of $(n+2)$ in the (2,2) entry of Eq.~\eqref{eq:5.1} 
    does not follow from Eq.~\eqref{eq:fdim6} and it is not predicted by constraints from mixing between the dimension-six singlet operators; we will comment on this factor in Sec.~\ref{sec:n=-2}.}
\begin{equation}\label{eq:5.1}
    \frac{d}{d\log\mu} \begin{pmatrix}
        c_{\phi^6}^{(S)}\\c_{\phi^4\square}^{(S)}
    \end{pmatrix}
    = 
    \begin{pmatrix}
        \gamma_{11} & (n-1)\,\gamma_{12}\\
        (n+2)(n+4)\,\gamma_{21} & (n+2)\gamma_{22}
    \end{pmatrix}
    \begin{pmatrix}
        c_{\phi^6}^{(S)}\\c_{\phi^4\square}^{(S)}
    \end{pmatrix}\,,
\end{equation}
where each $\gamma_{ij}$ is polynomial in $n$ and $\lambda$ (in a loop expansion for small $\lambda$). 
As noted in Sec.~\ref{sec:pertProof}, the off-diagonal factors $(n-1)$ and $(n+2)(n+4)$ are essential for the spectrum constraints to hold in the presence of operator mixing.
For instance, the $B_4$ representation contains a single dimension-six Lorentz scalar, $\mathcal{O}_{\phi^4\square}^{(B_4)}$, whereas the singlet representation contains two: $\mathcal{O}_{\phi^4\square}^{(S)}$ and $\mathcal{O}_{\phi^6}^{(S)}$. For the scaling dimensions of $\mathcal{O}_{\phi^4\square}^{(B_4)}$ and $\mathcal{O}_{\phi^4\square}^{(S)}$ to agree at $n=1$, the effect of mixing between the two singlet operators must vanish at this value of $n$. The factor $(n-1)$ ensures this, so that the relevant scaling dimension is determined by $\gamma_{22}$ alone.
In other words, the mixing of $c_{\phi^4\square}^{(S)}$ into $c_{\phi^6}^{(S)}$  vanishes as $n\to1$ since the $\phi^4\square$ operator becomes evanescent at $n=1$ while $c_{\phi^6}^{(S)}$ does not.

The same reasoning applies to the spectrum constraints relating the singlet operator to those in the $T_4$ and $T_6$ representations at $n=-2$ and $n=-4$, respectively. 
The $\phi^6$-type operator is evanescent at these values of $n$, while the $\phi^4\square$-type operator is not.
This implies the overall factor $(n+2)(n+4)$ in the $(2,1)$ entry of Eq.~\eqref{eq:5.1}. 
As a consequence of the term at order $n^2$, this entry can first become nonzero only at three loops. This non-renormalization result follows from an argument similar to that of Sec.~\ref{Sec:family}.
Incidentally, this one- and two-loop zero in the mixing between operators with distinct number of fields also follows from the non-renormalization rule of Ref.~\cite{Bern:2019wie}.

\paragraph{Dimension eight.}

At dimension eight, there are four singlet operators, which arise in the decomposition of the following four operators~\cite{Fonseca:2017lem,Fonseca:2019yya} (see also \cite[App.\,D]{Henriksson:2025vyi}),
\begin{align}
    &\phi^{a_1}
    \cdots \phi^{a_8}
    &&\leftrightarrow \ \,
    \raisebox{0.4mm}{\scalebox{0.35}{$\ydiagram{8}$}}_{\,S_8} \simeq \,
    \raisebox{0.4mm}{\scalebox{0.35}{$\ydiagram{8}$}}_{\,O(n)}
    +
    \raisebox{0.4mm}{\scalebox{0.35}{$\ydiagram{6}$}}_{\,O(n)}
    +
    \raisebox{0.4mm}{\scalebox{0.35}{$\ydiagram{4}$}}_{\,O(n)}
    +
    \raisebox{0.4mm}{\scalebox{0.35}{$\ydiagram{2}$}}_{\,O(n)}
    +
    S\,,\nn\\[2mm]
    &\phi^{a_1}\cdots\phi^{a_6}\square 
    &&\leftrightarrow \ \,
    \raisebox{1.2mm}{\scalebox{0.35}{$\ydiagram{4,2}$}}_{\,S_6} \simeq \,
    \raisebox{1.2mm}{\scalebox{0.35}{$\ydiagram{4,2}$}}_{\,O(n)}
    +
    \raisebox{1.2mm}{\scalebox{0.35}{$\ydiagram{2,2}$}}_{\,O(n)}
    +
    \raisebox{1.2mm}{\scalebox{0.35}{$\ydiagram{3,1}$}}_{\,O(n)}
    +
    \raisebox{0.4mm}{\scalebox{0.35}{$\ydiagram{4}$}}_{\,O(n)}
    +
    2\times \raisebox{0.4mm}{\scalebox{0.35}{$\ydiagram{2}$}}_{\,O(n)}
    +
    S\,,\nn\\[2mm]
    &\phi^a\phi^b\phi^c\phi^d \square^2 &&\leftrightarrow \ \,
    \raisebox{0.4mm}{\scalebox{0.35}{$\ydiagram{4}$}}_{\,S_4} \simeq \,
    \raisebox{0.4mm}{\scalebox{0.35}{$\ydiagram{4}$}}_{\,O(n)}
    +
    \raisebox{0.4mm}{\scalebox{0.35}{$\ydiagram{2}$}}_{\,O(n)}
    +
    S\,,\nn\\[2mm]
    &\phi^a\phi^b\phi^c\phi^d \square^2 &&\leftrightarrow \ \,
    \raisebox{1.2mm}{\scalebox{0.35}{$\ydiagram{2,2}$}}_{\,S_4} \hspace{4.2mm}\simeq \, 
    \raisebox{1.2mm}{\scalebox{0.35}{$\ydiagram{2,2}$}}_{\,O(n)}
    +
    \raisebox{0.4mm}{\scalebox{0.35}{$\ydiagram{2}$}}_{\,O(n)}
    +
    S\,.\label{eq:5.3}
\end{align}
The total dimensions of the reducible representations are 
\begin{align}
    &\text{dim}(\,\raisebox{0.4mm}{\scalebox{0.35}{$\ydiagram{8}$}}_{\,S_8}) = \binom{n+7}{8}\,,
    %
    \quad \text{dim}(\,\raisebox{1.2mm}{\scalebox{0.35}{$\ydiagram{4,2}$}}_{\,S_6}) 
    = \tfrac{1}{80} (n-1) n^2 (n+1) (n+2) (n+3)\,,\nn\\[2mm]
    &\text{dim}(\,\raisebox{0.4mm}{\scalebox{0.35}{$\ydiagram{4}$}}_{\,S_4}) = \tfrac{1}{24} n (n+1) (n+2) (n+3)\,,
    %
    \quad
    \text{dim}(\,\raisebox{1.2mm}{\scalebox{0.35}{$\ydiagram{2,2}$}}_{\,S_4}) = \tfrac{1}{12} (n-1) n^2 (n+1)\,, 
\end{align}
where by dimension we mean the number of independent components of the operators. This is the
result of the hook content formula, or the dimension of the corresponding $GL(n)$ representation. Restricting again to the singlet operators, it will be enough to note that these disappear from the operator spectrum whenever the corresponding operator with the general flavor structure has zero components. (Note that these do not have a negative number of components for any choice of $n$.) This results in the following structure in the mixing matrix,
\begin{align}
    \frac{d}{d\log\mu} 
    \pmatrixx{c^{(S)}_{\phi^8}\\[0.7mm]
    c^{(S)}_{\phi^6\square}\\[0.7mm]
    c^{(S)}_{\phi^4\square^2,1}\\[0.7mm]
    c^{(S)}_{\phi^4\square^2,2}}
=
\pmatrixx{\gamma_{11} & (n-1)\gamma_{12} & 
\gamma_{13} & (n-1)\gamma_{14}\\[2mm]
(n+6)(n+4)\gamma_{21} &\gamma_{22} &\gamma_{23} & \gamma_{24}\\[2mm]
(n+6)(n+4)\gamma_{31}&(n-1)\gamma_{32}&\gamma_{33}&(n-1)\gamma_{34}\\[2mm]
(n+6)(n+4)(n+2)\gamma_{41}&
    (n+2)\gamma_{42}&(n+2)\gamma_{43}
    &(n+2)\gamma_{44}}
\pmatrixx{c^{(S)}_{\phi^8}\\[0.7mm]
    c^{(S)}_{\phi^6\square}\\[0.7mm]
    c^{(S)}_{\phi^4\square^2,1}\\[0.7mm]
    c^{(S)}_{\phi^4\square^2,2}}
    \,,
    \label{eq:5.4}
\end{align}
which explains all patterns observed in the explicit results of Ref.~\cite{RoosmaleNepveu:2024zlz}, except for the factor of $(n+2)$ in the $(4,4)$ entry, on which we comment below.

Finally, we remark that the decomposition of
$\,\raisebox{1.2mm}{\scalebox{0.35}{$\ydiagram{4,2}$}}_{\,S_6}$ into $O(n)$ irreps contains the $T$ irrep twice; see Eq.~\eqref{eq:5.3}. For $n=0$, this irrep specializes as minus the singlet irrep. On the other hand, the $\,\raisebox{1.2mm}{\scalebox{0.35}{$\ydiagram{3,1}$}}_{\,O(n)}$ irrep in the same decomposition specializes as a singlet (with a plus sign) for $n=0$. This opens the intriguing possibility that the operators do not leave the spectrum in pairs, but in a group of four at the same time, resulting in more complicated spectrum constraints. We leave the study of this annihilation mechanism and the relevant scaling dimensions to future work.

\paragraph{Dimension ten $\phi^6\square^2$-type operators.}

Finally, at dimension ten, we restrict to the mixing matrix among operators with six fields and four derivatives. The operators with general flavor structure can be decomposed into seven operators belonging to six distinct irreducible representations of $S_6$~\cite{Fonseca:2017lem,Fonseca:2019yya},
\begin{align}
    \raisebox{0.4mm}{\scalebox{0.35}{$\ydiagram{6}$}}_{\,S_6}\,, \ 
    2\times\raisebox{1.2mm}{\scalebox{0.35}{$\ydiagram{4,2}$}}_{\,S_6}\,, \ 
    \raisebox{1.2mm}{\scalebox{0.35}{$\ydiagram{5,1}$}}_{\,S_6}\,, \ 
    \raisebox{2mm}{\scalebox{0.35}{$\ydiagram{3,2,1}$}}_{\,S_6}\,, \ 
    \raisebox{2mm}{\scalebox{0.35}{$\ydiagram{2,2,2}$}}_{\,S_6}\,, \ 
    \raisebox{3.6mm}{\scalebox{0.35}{$\ydiagram{2,1,1,1,1}$}}_{\,S_6}\,.
\end{align}
When decomposing these operators into irreducible $O(n)$ representations using the Littlewood restriction rule~\cite{10.1098/rsta.1944.0003,Littlewood1950}, only the first irrep, the second (which has two operators) and the fifth contain an $O(n)$ singlet. 
These singlets are evanescent whenever the dimensions of the operators vanish, leading to the mixing matrix
\begin{align}
\frac{d}{d\log\mu} \pmatrixx{c_{6,1}\\
    c_{6,2}\\c_{6,3}\\c_{\phi^4\square}}
=
\pmatrixx{\gamma_{11} & (n-1)\gamma_{12} & 
(n-1)\gamma_{13} & (n-1)(n-2)\gamma_{14}\\
(n+4)\gamma_{21} &\gamma_{22} &\gamma_{23} & (n-2)\gamma_{24}\\
(n+4)\gamma_{31}&\gamma_{32}&\gamma_{33}&(n-2)\gamma_{34}\\
(n+4)(n+2)\gamma_{41}&
    (n+2)\gamma_{42}&(n+2)\gamma_{43}
    &(n+2)\gamma_{44}}
\pmatrixx{c_{6,1}\\
    c_{6,2}\\c_{6,3}\\c_{\phi^4\square}}
    \,.\label{eq:5.6}
\end{align}
As was noted in Ref.~\cite{RoosmaleNepveu:2024zlz}, this mixing structure implies a one-loop zero in $\gamma_{14}$ and $\gamma_{41}$ because a degree two polynomial in $n$ is impossible with one-loop Feynman diagrams.
This non-renormalization result directly applies to the Standard Model EFT, noting that that a subset of pure-Higgs operators is invariant under $O(4)$ transformations. 
Unlike the argument in Sec.~\ref{Sec:family}, it does not extend to two loops, because the diagram of Eq.~\eqref{diag4.5} may have a $1/\varepsilon^1$ pole due to the presence of derivatives~\cite[Sec.\,5]{Jenkins:2023rtg}. In fact, the explicit results of~\cite{RoosmaleNepveu:2024zlz} show that $\gamma_{41}$ and $\gamma_{14}$ are non-zero at two loops.

\subsection[Non-renormalization for \texorpdfstring{$n=-2$}{n=-2}]{\boldmath Non-renormalization for $n=-2$}
\label{sec:n=-2}

The $(2,2)$ entry in Eq.~\eqref{eq:5.1}, as well as the $(4,4)$ entries in Eqs.~\eqref{eq:5.4} and~\eqref{eq:5.6} are proportional to $(n+2)$ to all orders. These diagonal entries cannot be explained through the arguments of the previous subsections, because they correspond to an operator renormalizing itself. 
Instead, these factors arise because the $\lambda\phi^4$ interaction, which is the $O(n)$ singlet component in the 
decomposition of $\raisebox{0.4mm}{\scalebox{0.35}{$\ydiagram{4}$}}_{\,S_4}$, vanishes at $n=-2$;
one cannot antisymmetrize four indices that take only two distinct values.
Said otherwise, the $
O(n)$ model becomes a free theory for $n=-2$. 
These zeros for $n=-2$ were previously explained through a diagrammatic argument~\cite{PhysRevLett.30.544}, and generalized to higher-point interactions in Ref.~\cite{Fisher:1973zzb}.
It was also previously noted that the $Sp(2)$ theory is a free theory~\cite{Fei:2015kta,Fraser-Taliente:2026iuj}.

Even though the theory is free for $n=-2$, we observe that not all anomalous dimensions vanish at this value of $n$. 
Importantly, only evanescent operators can have non-trivial anomalous dimensions at $n=-2$.
This includes any operator that branches from a $S_m$ representation with more than two columns. For negative $n$, these operators demand antisymmetrization of more than two indices.
In this paper, anomalous dimensions proportional to $(n+2)$ occur for the singlet operators branching from $\raisebox{0.4mm}{\scalebox{0.35}{$\ydiagram{2}$}}_{\,S_2}$,\,
$\raisebox{1mm}{\scalebox{0.35}{$\ydiagram{2,2}$}}_{\,S_4}\!$ and\,
$\raisebox{1.7mm}{\scalebox{0.35}{$\ydiagram{2,2,2}$}}_{\,S_6}$; the same is true of the field anomalous dimension $\gamma_\phi$~\cite{PhysRevLett.30.544}. These operators are genuine members of the spectrum at $n=-2$, but their scaling dimensions take their free-theory values.

In the plot of Fig.~\ref{fig:examplesDim6}, the vanishing of the anomalous dimension of one of the dimension-six singlet operators has been indicated by a red dot. This dot anchors the set of scaling dimensions of the $\phi^4\square$-type operators  to the $x$-axis, which are then interconnected through spectrum constraints at different values of $n$.

\section{Conclusion and Outlook}\label{sec:concl}

In this work, we studied the spectrum constraints that follow from degeneracies in the decomposition of tensor structures into irreducible representations of $O(n)$ for specific values of $n$~\cite{Cao:2023psi}. We extended them to even negative integers using the correspondence between $O(2N)$ and $Sp(-2N)$, and we made the mechanism behind these constraints more transparent
by working them out to all orders in perturbation theory in specific examples.
Our perturbative perspective allows us to pinpoint the operators whose scaling dimensions coincide when their representations are degenerate, which is relevant when there are multiple operators with the same quantum numbers. 
In principle, the work of Ref.~\cite{Cao:2023psi} and our arguments do not exclude more exotic spectrum constraints that involve more than two operators at the same time. We identified a set of four operators which may exhibit such an intricate annihilation mechanism below Eq.~\eqref{eq:5.4}, which would be worth studying in more detail.
In addition, while all our examples and arguments concern negative even values of $n$, it would be interesting to understand whether analogous spectrum constraints exist at negative odd values of $n$. Theories with a global $OSp(1|2N)$ symmetry, involving both commuting and anticommuting fields~\cite{McKane:1979rm}, would be a natural candidate to consider~\cite{Deligne}.

At leading orders in perturbation theory, we used spectrum constraints at fixed integer values of $n$ to identify degeneracies in the anomalous dimensions that persist for arbitrary $n$.
This insight makes it possible to `bootstrap' the scaling dimensions at low orders in perturbation theory, which we exemplify by deriving a novel closed-form two-loop formula for the scaling dimensions of 
$\phi^k$\nobreakdash-type operators in any representation of $O(n)$, 
taking as input only the known scaling dimension of two operators of this type.
In our approach, we completely determine the two-loop scaling dimensions of all $\phi^k$-type operators in terms of two functions, which are fixed from existing results in the literature.

Finally, although the spectrum constraints relate scaling dimensions of operators in different representations, we showed that the same mechanism also leaves an imprint on the renormalization-group mixing matrix within a single representation. In an appropriate operator basis, identified using a general EFT approach, certain matrix elements must contain overall factors of $(n\pm a)$.
When such factors are diagrammatically impossible, they imply non-renormalization results at low loop orders, cf.\ Eq.~\eqref{eq:5.6}. 
It would be interesting to uncover analogous non-renormalization results in more complicated theories, such as the Standard Model EFT, by expressing their renormalization group equations in the appropriate operator basis.

There are various possible directions for future research, of which we highlight two in more detail: constraints on operator product expansion (OPE) coefficients and constraints from spacetime evanescence when treating $d$ as a continuous variable.

\subsection{Operator product expansion}
Besides constraining the operator spectrum, the evanescence of operators for specific integer values of $n$ should also impact other observables, such as OPE coefficients. These constraints are straightforward to derive in perturbation theory by decomposing general tensor structures into irreducible representations.
Consider, for instance, the OPE of $c^a \phi^a(x)\, c^b \phi^b(0)$ in the $O(n)$ model, which takes the schematic form%
    \footnote{In general, the OPE takes this simple form only at the fixed point.}
\begin{align}
    \langle c\cdot \phi(x) \, 
            c\cdot \phi(0) \, \cdots \rangle 
    &= \sum_\mathcal{O} \frac{C_{\phi\phi\mathcal{O}}^{ab}(x)}{|x|^{2\Delta_\phi - \Delta_\mathcal{O}}} 
    \langle \mathcal{O}^{ab}(0) \cdots \rangle 
    = \frac{c^ac^b f_1 + \delta^{ab}c^2 f_2}{|x|^{2\Delta_\phi - \Delta_{\phi^2}}}
    \langle \mathcal{O}_{\phi^2}^{ab}(0) \cdots \rangle  + ...\,,
\end{align}
where $f_1$ and $f_2$ are polynomials in the coupling $\lambda$ and the number of flavors $n$. We suppress terms that are subleading for small $x$. 
Similar to the derivation of the spectrum constraints, the product tensor structure can be decomposed as $c^a c^b = \left(c^a c^b - \tfrac{c^2}{n}\delta^{ab}\right) + \tfrac{c^2}{n}\delta^{ab}$. 
If we define the singlet OPE coefficient as the coefficient of $\delta^{ab}c^2/n$, and the traceless symmetric OPE coefficient as that of $(c^ac^b-\frac{c^2}{n}\delta^{ab})$,
we obtain
\begin{align}
C_{\phi \phi \phi^2_S} = f_1 + n f_2\,, &&
C_{\phi \phi \phi^2_T} = f_1\,,
\end{align}
showing in full generality that these agree for $n=0$, as do the scaling dimensions of the singlet and traceless symmetric operators.
Indeed, 
in the critical $O(n)$ model in $d=4-\eps$ dimensions, these OPE coefficients are known up to order $\eps^3$~\cite{Dey:2016mcs} (see also~\cite{Henriksson:2022rnm}), 
\begin{align}
C^2_{\phi \phi \phi^2_S} &= 
2-(2+n)\left( 
    \frac{2}{n+8}\eps 
    +
    \frac{6\,(14+3n)}{(n+8)^3}\eps^2
    +O(\eps^3)
\right),\nn\\
C^2_{\phi \phi \phi^2_T} &= 
2-\frac{4}{n+8}\eps 
-\frac{12\,(14+3n)}{(n+8)^3}\eps^2 + O(\eps^3)\,,
\label{6.3}
\end{align}
confirming that $C^2_{\phi \phi \phi^2_S}|_{n=0} = 
C^2_{\phi \phi \phi^2_T}|_{n=0}$.
To expose this relation, we changed the overall normalization convention w.r.t.\ Ref.~\cite{Dey:2016mcs} and we present the result only up to order $\eps^2$ for brevity. 
In the normalization convention of Ref.~\cite{Henriksson:2022rnm},  $C^2_{\phi \phi \phi^2_T}$ is proportional to $\sqrt{(n-1)(n+2)}$, which ensures that the OPE coefficient vanishes whenever the dimension of the $T$ representation vanishes.

We also remark on the overall factor of $(n+2)$ in 
the loop correction to $C_{\phi \phi \phi^2_S}$ and its absence in
$C_{\phi \phi \phi^2_T}$. As explained in Sec.~\ref{sec:n=-2}, this can be explained from the fact that 
the singlet $\phi^2_S$ operator exists in the $Sp(2)$ model, while the marginal $\phi^4$ interaction vanishes. On the other hand, $\phi^2_T$ is evanescent for $n=-2$ and its OPE coefficient  is not constrained in this way.

It will be interesting to study the constraints from evanescence on the OPE coefficients in more detail, and whether they can be leveraged to infer new data in perturbation theory without additional computations, as we exemplified for the spectrum constraints in Sections \ref{Sec:family} and \ref{sec:3}. 
For this purpose, it will prove useful to systematically study the OPE in scalar $\phi^4$ theories with general flavor structure.

\subsection{Spacetime evanescence}
We have considered spectrum constraints implied by the representation theory of the global $O(n)$ symmetry group when treating $n$ as a variable. If we instead continue the number of spacetime dimensions to any $d$, similar spectrum constraints may be expected due to evanescence at
 specific values of $d$~\cite{Zan:2026oyb}.
As the simplest example, we could apply the logic leading to Eq.~\eqref{2.5} to an operator $c^{\mu\nu}\mathcal{O}^{\mu\nu}$. Upon reducing the Lorentz structure to irreducible representations, this leads to a (naive) prediction 
of spectrum constraints between
scalar operators and operators in the spin-two Lorentz representation for $d=0$. 

The $\varepsilon$ expansion is not well suited to check such spectrum constraints, because the resulting expressions are not exact functions of $d$.
Other potential obstacles include renormalization scheme ambiguities due to the possible appearance of 
    $\varepsilon/\varepsilon$ terms after taking traces
    in counterterms for general Lorentz representations, as well as operator basis ambiguities due to the equation of motion when derivatives are contracted.
For this reason, other methods that are exact in $d$ will prove more insightful. 
Indeed, expanding the large-$n$ result for the scaling dimensions of the operators of the schematic form $\mathcal{O}_{S,0} \sim \sigma^2 \square$ and 
$\mathcal{O}_{S,2} \sim \sigma^2 \partial^2$ 
(both in the $O(n)$ singlet representation) for small $d=0+\alpha$, we find~\cite{Vasiliev:1993ux,Lang:1992zw,Henriksson:2022rnm}:
\begin{align}
    \Delta_{S,0} &= 6-\left(\frac{7\,\alpha}{n} - \frac{41\,\alpha^2}{12\,n} + O(\alpha^3)\right) + O(1/n^2)\,,\nn\\
    \Delta_{S,2} &=  6-\left(\frac{7\,\alpha}{n} - \frac{5\,\alpha^2}{12\,n} + O(\alpha^3)\right) + O(1/n^2)\,.
\end{align}
Remarkably, the two dimensions agree not only at $d=0$, but through first order in $\alpha$.
It would be useful to study such relations in more detail 
at large-$n$ or using other methods that treat the number of spacetime dimensions as a variable.

\section*{Acknowledgments}

We thank  W.~Cao, T.~Melia and L.~Lindwasser for useful discussions and comments on the draft.
DA thanks R.~Klabbers for early discussions on the topic. 
DA is supported by the MUR PRIN contract 2022N9CTAE ”Constraining strongly coupled quantum field theories using symmetry”.
JRN is supported by the Yushan Young Scholarship 112V1039 from the Ministry of Education (MOE) of Taiwan, and also by the National Science and Technology Council (NSTC) grants 113-2112-M-002-038-MY3 and 114-2923-M-002-011-MY5.
During the writing stage of this work, we made use of large language models from OpenAI and Anthropic.

\bibliographystyle{JHEP}
\bibliography{refs}
 
\end{document}